\documentclass[review]{elsarticle}

\usepackage{lineno,hyperref}
\modulolinenumbers[5]

\usepackage[nolist]{acronym} 			
\usepackage{booktabs}					
\usepackage{tabularx}					
\usepackage{amsmath, amssymb,amsthm} 	
\newcommand{\ra}[1]{\renewcommand{\arraystretch}{#1}}	
\usepackage{subfig}						
\usepackage{scrextend}  				
\usepackage{tikz}						
\usetikzlibrary{trees}					
\usepackage{comment}					
\usepackage[official]{eurosym}			
\usepackage[at]{easylist}               

\usepackage{color,soul}

\usepackage{comment}                    
\usepackage{amsmath}
\usepackage{accents}
\newcommand{\ubar}[1]{\underaccent{\bar}{#1}}
\usepackage[long]{optidef}
\usepackage{lscape}
\usepackage{subfig}
\usepackage[official]{eurosym}
\usepackage{url}

\makeatletter
\newcounter{manualsubequation}
\renewcommand{\themanualsubequation}{\alph{manualsubequation}}
\newcommand{\startsubequation}{%
  \setcounter{manualsubequation}{0}%
  \refstepcounter{equation}\ltx@label{manualsubeq\theequation}%
  \xdef\labelfor@subeq{manualsubeq\theequation}%
}
\newcommand{\tagsubequation}{%
  \stepcounter{manualsubequation}%
  \tag{\ref{\labelfor@subeq}\themanualsubequation}%
}
\let\subequationlabel\ltx@label
\makeatother

\journal{arXiv}

\biboptions{sort&compress}

\usepackage{anysize}
\marginsize{2.5cm}{2.4cm}{1.5cm}{2cm}

\begin{document}

\begin{frontmatter}

\title{Renewable Energy Targets and Unintended Storage Cycling: \\
Implications for Energy Modeling}



\author[A1]{Martin Kittel\corref{MK}}
\ead{mkittel@diw.de}

\author[A1]{Wolf-Peter Schill\corref{WPS}}
\ead{wschill@diw.de}

\cortext[MK] {Corresponding author}

\address[A1] {DIW Berlin, Department of Energy, Transportation, Environment, Mohrenstra{\ss}e 58, 10117 Berlin, Germany}




\begin{abstract}

To decarbonize the economy, many governments have set targets for the use of renewable energy sources. These are often formulated as relative shares of electricity demand or supply. Implementing respective constraints in energy models is a surprisingly delicate issue. They may cause a modeling artifact of excessive electricity storage use. We introduce this phenomenon as ``unintended storage cycling'', which can be detected in case of simultaneous storage charging and discharging. In this paper, we provide an analytical representation of different approaches for implementing minimum renewable share constraints in energy models, and show how these may lead to unintended storage cycling. Using a parsimonious optimization model, we quantify related distortions of optimal dispatch and investment decisions as well as market prices, and identify important drivers of the phenomenon. Finally, we provide recommendations on how to avoid the distorting effects of unintended storage cycling in energy modeling.

\vspace{.5cm}


\begin{acronym}
 \acro{APC}{Across-Period-Cycling}
 \acro{BEV}{battery-electric vehicles}
 \acro{DIETER}{Dispatch and Investment Evaluation Tool with Endogenous Renewables}
 \acro{KKT}{Karush-Kuhn-Tucker}
 \acro{LCOE}{Levelized Cost of Electricity}
 \acro{LCOS}{Levelized Cost of Storage}
 \acro{MV}{Market Value}
 \acro{NSL}{Normalized Storage Conversion Losses}
 \acro{PV}{photovoltaics}
 \acro{RLDC}{residual load duration curve}
 \acro{SLCR}{Storage Loss Coverage by Renewables}
 \acro{SPC}{Same-Period-Cycling}
 \acro{$STO^{LOSS}$}{storage conversion losses}
 \acro{VRE}{Variable renewable energy}
\end{acronym}

\end{abstract}


\end{frontmatter}

\newpage

\section{Introduction}

Replacing conventional electricity generation with renewable energy sources is a prime option for mitigating greenhouse gas emissions in the power sector. Technological progress, economies of scale, and support measures have led to substantially growing shares of renewables in power sectors around the world. Firm renewable capacity potentials, i.e.,~for hydroelectric and biomass power, are limited in many regions. \ac{VRE}, such as wind and solar \ac{PV}, will thus play a major role in the further transition. These technologies have variable generation characteristics, as their temporal availability depends on weather conditions, such as wind speeds and cloud cover \cite{joskow_comparing_2011}. Investment costs for \ac{VRE} have tremendously decreased in recent years \cite{iea_renewables_2020}. Due to an insufficient internalization of greenhouse gas emissions \cite{ricke_country-level_2018} and low marginal generation costs of existing thermal power plants, policy support continues to be relevant for low-cost renewable deployment above penetration levels that would emerge without policy intervention \cite{polzin_how_2019}.

Many governments on state, national, and even intergovernmental levels have adopted a minimum renewable energy share by a target date to promote the deployment of renewable energy \cite{ozdemir_capacity_2020}. Essentially, this share describes the ratio of energy from renewable sources to the total energy supplied or consumed in a specific economic sector, or in the entire energy system. Such targets and related policy measures can facilitate investment in \ac{VRE} technologies, and usually rise over time.

For instance, Germany has a long tradition of setting renewable energy targets for its power sector, in particular via the Renewable Energy Sources Act (EEG). The 2021 version of the EEG includes a renewable share of 65\% in gross electricity consumption by 2030 \cite{bundesregierung_gesetz_2020}.\footnote{Gross electricity consumption includes end energy consumption, grid and storage losses, internal power plant consumption required for operation, and electricity exports net of imports.} France and Spain aim for a share of 40\% and 74\% of total generation by 2030, respectively, while Sweden envisages a fully renewable power sector by 2040 \cite{ren21_renewables_2021}. In the United States, renewable portfolio standards require a certain share of a utility's total generation to come from renewable energy sources. In 2020, 30 states had adopted mandatory standards, while in eight states voluntary standards are in place~\cite{zhou_renewable_2020}. For instance, California aims at a 60\% (100\%) renewable share by 2030 (2045)~\cite{california_puc_californias_2021}. China targets a renewable share of 35\% in its total electricity generation by 2030 \cite{ren21_renewables_2020}. Japan aims at 22-24\% for the same target year. Overall, 137 countries had renewable energy targets in place for the power sector in 2020 \cite{ren21_renewables_2021}. While some of these countries have set only absolute renewable power capacity targets (e.g., Switzerland or New Zealand), most make use of some sort of minimum renewable energy target. 

\begin{figure}[htbp]
\centering
\noindent\includegraphics[width=\linewidth,height=\textheight, keepaspectratio]{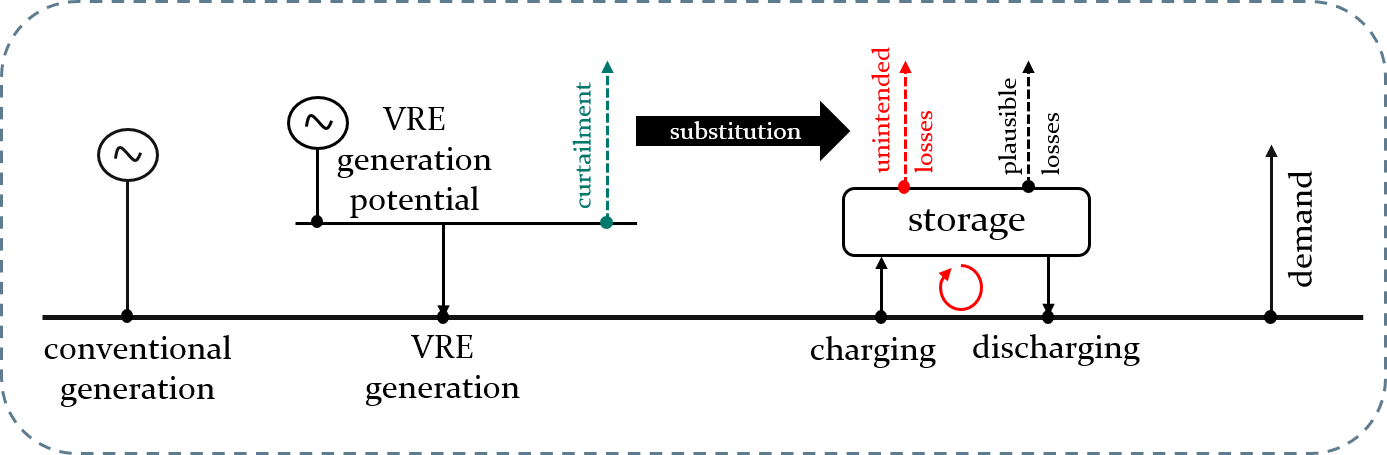}
\caption{Stylized illustration of unintended storage cycling: Instead of curtailing (dashed green) \ac{VRE} surplus, the optimization may lead to unintended storage cycling, i.e.,~to additional storage conversion losses (dashed red), which remove excess renewable electricity from the system.}
\label{fig:mechanism_sto_cyc}
\end{figure}

Capacity expansion models of the power sector or the entire energy system are frequently used to explore future scenarios with high shares of renewables and determine least-cost solutions \cite{dagoumas_review_2019}. To comply with climate policy, such models are often constrained by renewable energy targets. In this case, an unexpected modeling artifact related to \ac{VRE} generation may occur. Figure \ref{fig:mechanism_sto_cyc} illustrates the general mechanism. At high shares of \ac{VRE}, situations arise in which the \ac{VRE} generation potential temporarily exceeds the demand for electricity. In these situations, curtailing the renewable surplus would be an obvious option (dashed green arrow). However, certain minimal renewable share constraint formulations may lead to a situation where renewable curtailment is replaced by storage conversion losses (red dashed error), facilitated by additional, and often simultaneous, storage charging and discharging. Such \textit{unintended storage cycling} drives up generation from \ac{VRE} compared to the first-mentioned option of curtailing renewable surplus generation, and may thus facilitate achieving renewable energy constraints at lower costs. Note that this mechanism may arise in energy models, but does not relate to real-world storage applications.

The phenomenon of unintended storage cycling has, to the best of our knowledge, not been described or discussed in the energy system literature \cite{bistline2020}. There is reason to believe that this artifact is undetected or ignored, yet prevailing in the power sector and energy system modeling domain. Our paper aims at closing this gap. To this end, we first provide a systematic representation of different approaches for implementing minimum renewable share constraints in energy models. Using a parsimonious optimization model of a power sector, we analytically derive some intuition as to how these constraints may lead to unintended storage cycling and examine their impact on equilibrium conditions of storage operators. For a stylized model parameterization, we numerically analyze how this phenomenon may distort model outcomes. We quantify the effects on dispatch and investment decisions of both storage and other technologies, and on market prices. We also explore important drivers of unintended storage cycling. Finally, we discuss different options for avoiding the distorting effects of unintended storage cycling in energy modeling and provide recommendations for energy modelers.

In energy modeling applications, constraints requiring a certain renewable share may be manifold. A review by Gillich et al.~\cite{gillich_impacts_2019} shows that some applications use constraints for installed renewable generation capacity (e.g.,~in GW). 
Other model applications use constraints on the share of renewable energy (e.g.,~in TWh). Table~\ref{tab:overview_models} shows a selection of well-established European models and model applications that employ renewable capacity and energy targets. Additionally, two commonly implemented climate policy instruments are listed, which also influence the share of renewable energy sources in model outcomes: an annual green house gas emission budget and an explicit CO$_2$ price. The information is based on a review of research articles, model documentations as well as other references such as dissertations, and, wherever possible, on personal validation by model developers or affiliated modelers. 

Renewable energy constraints in power sector and energy system models may come in various forms. Energy provision from renewable (conventional) energy sources has to achieve (must not exceed) a certain share in total demand or generation within a particular sector, or in the entire energy system. Additionally, storage losses may be covered to a different extent in such constraints. Based on this survey, we systematize possible ways of implementing renewable energy targets, and examine their impact on modeling results.

The artifact of unintended storage cycling potentially can occur in all cost-optimizing power sector and energy system models that include binding renewable targets. Personal communication further confirms that the phenomenon has already been observed in various European energy models, which are marked with an asterisk (*) in Table~\ref{tab:overview_models}. For entries without an asterisk, it was not possible to clarify whether the artifact has been observed so far.

Not all of the listed models consider storage losses in the formulation of the renewable target, and only a minority does so to the full extent. We discuss how this impacts the results of linear optimization models. As we show in this paper, an incomplete consideration of storage losses in the renewable energy constraint causes unintended storage cycling. To the best of our knowledge, this has not been discussed in the energy modeling literature before.

The remainder of the paper is structured as follows: Section \ref{sec:methods} introduces a parsimonious power sector model and a numerical model implementation. Section \ref{sec:analytical_part} systematizes possible renewable energy target constraint formulations, provides intuition for the cause of unintended storage cycling, and discusses the artifact's implications for storage in the long-term equilibrium. Section \ref{sec:usc_types} illustrates different types of unintended storage cycling. Section \ref{sec:numerical_results} contains a numerical analysis, and determines the artifact's impact on dispatch and investment decisions, storage system effects, prices, and its drivers. Section \ref{sec:discussion} discusses other instances of and potential remedies for unintended storage cycling. Section \ref{sec:summary_conclusion} summarizes and concludes. Additional material is available in the supplementary information (SI).

\begin{landscape}
\begin{table}[htbp]
\centering
\ra{1.1}
\caption{Model features that affect the shares of renewable energy sources in selected European energy models and model applications.}
\label{tab:overview_models}
\vspace{.1cm}
\scriptsize
\begin{tabular*}{1.1\textwidth}{@{\extracolsep{\fill}}lcccccccr@{}}\toprule
        & renewable & conventional & renewable & conventional & renewable & CO$_2$ & CO$_2$ & docu-\\
Model   & \multicolumn{2}{c}{share in demand} & \multicolumn{2}{c}{share in generation} & capacity target & budget & price & mentation \\
\midrule
AnyMOD   & & & & & & x$^a$ & x$^a$ & \cite{goke_graph-based_2021} \\
Calliope* & x\textsuperscript{\tiny\cite{trondle_tim_euro-calliope_2020}}$^b$ & x$^a$ & x$^a$ & x$^a$ & x\textsuperscript{\tiny\cite{pfenninger_renewables_2015,pfenninger_dealing_2017}} & x\textsuperscript{\tiny\cite{pickering_quantifying_2021}} & x\textsuperscript{\tiny\cite{pickering_quantifying_2021}} & \cite{pfenninger_calliope_2018} \\
DIETER*   & & x\textsuperscript{\tiny\cite{zerrahn2017,schill2017,schill_long-run_2018}} & x\textsuperscript{\tiny\cite{schill_flexible_2020,stockl_optimal_2021}} & & & x$^b$ & x$^b$ & \cite{zerrahn2017,gaete-morales_dieterpy_2021} \\
DIMENSION* & x$^c$ & x$^c$ & x$^c$ & x$^c$ & x\textsuperscript{\tiny\cite{brundlinger_dena-leitstudie_2018}} & x\textsuperscript{\tiny\cite{helgeson_role_2020}} & x$^a$ & \cite{richter_dimension_2011,helgeson_role_2020} \\
dynELMOD* & & & & & x\textsuperscript{\tiny\cite{kittel_scenarios_2020}} & x\textsuperscript{\tiny\cite{gerbaulet_european_2019}} & & \cite{gerbaulet_dynelmod_2017} \\
EMMA*    & x$^b$ & & & & x\textsuperscript{\tiny\cite{hirth_market_2013,ruhnau_heating_2020,ruhnau_how_2021,ruhnau_why_2021}} & x$^a$ & x\textsuperscript{\tiny\cite{hirth_market_2013,ruhnau_heating_2020,ruhnau_how_2021,ruhnau_why_2021}} & \cite{hirth_market_2013} \\
ELTRAMOD & & & & & x\textsuperscript{\cite{anke_expansion_2021}} & x\textsuperscript{\cite{anke_coal_2020,hobbie_windfall_2019}} & x\textsuperscript{\cite{eising_future_2020}} & \cite{ladwig_demand_2018} \\
EU-REGEN* & & & x$^a$ & & & x\textsuperscript{\tiny\cite{mier_power_2020,weissbart_decarbonization_2020,azarova_market_2021}} & x\textsuperscript{\tiny\cite{mier_unraveling_2021,siala_which_2020}} & \cite{weissbart_framework_2019} \\
E2M2     & x\textsuperscript{\cite{fleischer_systemeffekte_2019}} & & x\textsuperscript{\cite{torralba-diaz_identification_2020}} & & x\textsuperscript{\cite{schick_role_2020}} & x\textsuperscript{\tiny\cite{gillich_impacts_2019,fleischer_systemeffekte_2019,guthoff_quantification_2021,torralba-diaz_identification_2020}} & x\textsuperscript{\cite{torralba-diaz_identification_2020,schick_role_2020}} & \cite{fleischer_systemeffekte_2019} \\
ENTIGRIS* & x\textsuperscript{\tiny\cite{senkpiel_autgrid_2019}} & x\textsuperscript{\tiny\cite{senkpiel_autgrid_2019}} & & & & & x\textsuperscript{\tiny\cite{senkpiel_autgrid_2019}} & n/a$^d$ \\
GENESYS  & & & & & & x$^c$ & x$^c$ & \cite{bussar_optimal_2014,bussar_investigation_2019} \\
GENeSYS-MOD* & & & x\textsuperscript{\tiny\cite{loeffler_designing_2017}} & & & x\textsuperscript{\tiny\cite{hainsch_emission_2021}} & x\textsuperscript{\tiny\cite{burandt_genesys-mod_2018}} & \cite{burandt_genesys-mod_2018}\\
ISAaR    & & & & & & x$^a$ & x$^c$ & \cite{boing_hourly_2019}  \\
LIMES-EU* & x\textsuperscript{\tiny\cite{pietzcker_tightening_2021}} & & x$^a$ & & & x\textsuperscript{\tiny\cite{pietzcker_tightening_2021}} & x\textsuperscript{\tiny\cite{osorio_how_2020}} & \cite{osorio_documentation_2020,knopf_european_2015} \\
LUT Model* & x$^b$ & x\textsuperscript{\tiny\cite{bogdanov_low-cost_2021}} & x\textsuperscript{\tiny\cite{solarpower_europe_100_2020}}\ & x\textsuperscript{\tiny\cite{solarpower_europe_100_2020}} & x\textsuperscript{\tiny\cite{renewable_energy_institute_renewable_2021}} & x$^a$ & x\textsuperscript{\tiny\cite{satymov_value_2021}} & \cite{bogdanov_low-cost_2021}\\
oemof & x$^c$ & x$^c$ & & & x\textsuperscript{\cite{hilpert_analysis_2020}} & x$^c$ & x$^c$ & \cite{hilpert_analysis_2020} \\
PyPSA* & x$^c$ & & x$^c$ & & x$^c$ & x\textsuperscript{\cite{parzen_beyond_2021}} & x\textsuperscript{\cite{neumann_near-optimal_2021}} & \cite{brown_pypsa_2018}\\
REMix* & x\textsuperscript{\cite{xiao2020}} & & x\textsuperscript{\cite{gils_integrated_2017,scholz_application_2017}} & & & x\textsuperscript{\cite{moser2020}} & x\textsuperscript{\cite{gils2019}} & \cite{gils_integrated_2017,scholz_application_2017} \\ 
\bottomrule
\multicolumn{9}{l}{\begin{tabularx}{1.0\textwidth}{X}\textit{Notes:} * Personal communication confirms that the artifact has been observed in this model. $^a$~Possible, but not used or documented in model applications so far. $^b$~Used in model applications, but not published yet. $^c$~Applicable according to personal communication, but insufficient information for linking specific model applications to this particular constraint. $^d$~More information available at \url{www.ise.fraunhofer.de/de/geschaeftsfelder/leistungselektronik-netze-und-intelligente-systeme/energiesystemanalyse/energiesystemmodelle-am-fraunhofer-ise/entigris.html}.
\end{tabularx}}	\\
\end{tabular*}
\end{table}
\end{landscape}

\section{Methods}\label{sec:methods}

\subsection{Formal definition of a parsimonious dispatch and investment model}\label{ssec:model_definition}

We use a parsimonious power sector optimization model to analytically show how different minimum renewable constraint formulations relate to costs and market values of storage. Adopting a long-run equilibrium perspective, the model minimizes total system costs of a full year. These include annualized investment and annual variable costs of generation and storage technologies. The model covers a single region and abstracts from a spatial resolution and grid congestion. We assume linear cost functions and price-inelastic demand, perfect foresight, and a perfect energy-only market. This set of assumptions renders a long-term market equilibrium with zero profits for market participants. The solution coincides with the optimal allocation determined by a benevolent social planer. The model set-up and derivations are similar to Brown~\&~Reichenberg~\cite{brown_decreasing_2021} and Biggar~\&~Hasemzadeh~\cite[Ch.4]{biggar_economics_2014}. Throughout the exposition, capital letters denote endogenous decision variables. Small letters refer to sets and exogenous parameters.

We denote $\sum_t d_t$ as the total demand for electricity over all hours $t$ of the year. The set of generators consists of firm and variable renewable generators $s \in \mathcal{R}$ and firm conventional generators $s \in \mathcal{C}$, while an unspecified $s$ applies to the set of all generators. $G_{s\in \mathcal{R},t}$ represents hourly renewable energy generation net of curtailment of renewable surplus $CU_{s,t}$. Equivalently, $G_{s\in \mathcal{C},t}$ is the hourly dispatch of conventional generators. Total annual electricity generation is $\sum_{s,t} G_{s,t} = \sum_{s \in \mathcal{R}, t} G_{s,t} + \sum_{s\in \mathcal{C}, t} G_{s,t}$. The parameter $\bar{g}_{s,t} \in [0,1]$ renders a time-variant availability of generator $s$. We denote a storage unit $r$, its charging from the grid $G^{in}_{r,t}$, and its discharging back to the grid $G^{out}_{r,t}$. Total storage loss is $\sum_{r,t} (G^{in}_{r,t} - G^{out}_{r,t})$, which has to be covered by either renewable or conventional generators. We denote $i_s$ and $i_r$ as technology-specific investment costs, $o_s$ and $o_r$ as operation and maintenance costs accruing for hourly dispatch. $C_s$, $C_r^{in}$ and $C_r^{out}$ denote installed capacity (power rating), and $C_r^l$ storage energy capacity. The hourly storage energy level is $G_{r,t}^l$. 

For the sake of readability, we define $\circ = \{l, in, out\}$ and $* = \{in, out\}$, and use the following aggregate cost terms for storage investment and usage in the objective function.

\begin{equation}\label{eq:investment_cost}
    \sum_{r,\circ} i_r^\circ C_r^\circ = \sum_r i_r^l C_r^l + \sum_r i_r^{in} C_r^{in} + \sum_r i_r^{out} C_r^{out}
\end{equation}

\begin{equation}\label{eq:variable_cost}
    \sum_{r,t,*} o_r^* G_{r,t}^* = \sum_{r,t} o_r^{in} G_{r,t}^{in} + \sum_{r,t} o_r^{out} G_{r,t}^{out}
\end{equation}
\vspace{0.1cm}

Investment costs (\ref{eq:investment_cost}) comprise annualized infrastructure cost of the storage (dis)charge unit, and storage energy costs. Storage usage cost (\ref{eq:variable_cost}) are variable and accrue whenever a storage unit (dis)charges. We formulate the optimization problem of the stylized power sector as follows:

\begin{mini!}|l|
    {\substack{\displaystyle C_s, G_{s,t},\\ \displaystyle C_r^\circ, G_{r,t}^\circ}}{\sum_s i_s C_s + \sum_{s,t} o_s G_{s,t} + \sum_{r,\circ} i_r^\circ C_r^\circ + \sum_{r,t,*} o_r^* G_{r,t}^*\label{eq:obj}}
    {}{}
	\addConstraint{-d_t + \sum_s G_{s,t} + \sum_r G_{r,t}^{out} - \sum_r G_{r,t}^{in}}{=0 \perp \lambda_t\label{eq:energy_balance}}{\quad \forall t}
	\addConstraint{G_{s,t}}{ \geq 0 \perp \ubar{\mu}_{s,t}\label{eq:nnc_s}}{\quad \forall s, t}
	\addConstraint{\bar{g}_{st} C_s - G_{s,t} - CU_{s,t}}{= 0 \perp \lambda^{CU}_{s,t}\label{eq:cu_s}}{\quad \forall s \in \mathcal{R}, t}
	\addConstraint{C_s - G_{s,t}}{\geq 0 \perp \bar{\mu}_{s,t}\label{eq:upper_s}}{\quad \forall s \in \mathcal{C}, t}
	\addConstraint{G^\circ_{r,t}}{\geq 0 \perp \ubar{\mu}^\circ_{r,t}\label{eq:nnc_r}}{\quad \forall r, t}
	\addConstraint{C^\circ_r - G^\circ_{r,t}}{\geq 0 \perp \bar{\mu}^\circ_{r,t}\label{eq:upper_r}}{\quad \forall s, t}
	\addConstraint{G^l_{r,t} - G^l_{r,t-1} - \eta^{in}_r G^{in}_{r,t} + (\eta^{out}_r)^{-1} G^{out}_{r,t}}{= 0 \perp \lambda^l_{r,t}\label{eq:sto_level}}{\quad \forall r, t}
	\addConstraint{\sum_{s \in \mathcal{R},t} G_{s,t} - \Theta}{\geq 0 \perp \mu_{\theta}\label{eq:theta}}{}
	\addConstraint{\Omega - \sum_{s \in \mathcal{C},t} G_{s,t}}{\geq 0 \perp \mu_{\omega}\label{eq:omega}}{}
\end{mini!}

The objective function (\ref{eq:obj}) minimizes total fixed and variable costs of all technologies. The market clearing condition (\ref{eq:energy_balance}) balances supply and demand for electricity at all times. Constraints (\ref{eq:nnc_s}) and (\ref{eq:nnc_r}) impose non-negativity. Variable renewable generation is limited by \ac{VRE} availability, and curtailment of \ac{VRE} surplus (\ref{eq:cu_s}), which comes at no costs. Generation of conventional generators may not exceed installed capacity (\ref{eq:upper_s}).  Storage (dis)charge and levels may be no larger than installed capacities (\ref{eq:upper_r}). Constraint (\ref{eq:sto_level}) ensures inter-temporal consistency of storage filling and withdrawal. As storage incurs conversion losses, storage (dis)charge effects on the storage level are corrected by the efficiency rates $\eta^{in}_r < 1$ and $\eta^{out}_r < 1$ (\ref{eq:sto_level}). 
To avoid free lunch, we further demand that the storage level in the first and last hour of the year need to be equal $G^l_{r,0} = G^l_{r,T}$. Constraint (\ref{eq:theta}) imposes a minimum renewable share, requiring total renewable generation aggregated overall $s \in \mathcal{R}$ to achieve at least $\Theta$. In contrast, constraint (\ref{eq:omega}) imposes a maximum share $\Omega$ on total conventional generation aggregated over all $s \in \mathcal{C}$. Constraints (\ref{eq:omega}) and (\ref{eq:theta}) are mutually exclusive substitutes. Only one is considered by the optimization problem. Table~\ref{tab:res_shares_opt} in the supplementary information provides unequivocal constraint formulations for our 12 model specifications aligned with the notation above.

A constraint's dual variable indicates how much the objective value changes if the constraint relaxed, also called shadow price. $\lambda_t$ represents the system's marginal cost of meeting an additional unit of demand for electricity. $\lambda^{CU}_{s,t}$ represents the change in total cost when marginally changing $CU_{s,t}$. In the optimum, it takes the value of zero. The dual variables $\ubar{\mu}_{s,t} \geq 0$, $\bar{\mu}_{s,t} \geq 0$ and $\ubar{\mu}^\circ_{r,t} \geq 0$, $\bar{\mu}^\circ_{r,t} \geq 0$, $\lambda^l_{r,t}$ are the shadow prices of the capacity and generation constraints of generators and storages, respectively. $\mu_{\theta / \omega} \geq 0$ indicate how much total system cost would increase, if the boundary of the binding constraint was relaxed. That is, if one more (less) unit of electricity from renewables (fossil fuels) had to be supplied. Loosely speaking, it is the marginal cost of tightening the \ac{VRE} policy.

\subsection{Numerical implementation}\label{ssec:numerical_model_defintion}

In section~\ref{sec:numerical_results}, we use a numerical implementation of our power sector model to derive optimal dispatch and capacity expansion decisions. It is a stylized version of the larger \ac{DIETER} capacity expansion model \cite{zerrahn2017}. A similar model set-up has been previously used for analyzing electricity storage needs for renewable energy integration \cite{zerrahn2018}, for reflections on the changing role of electricity storage in the renewable energy transition \cite{schill_electricity_2020}, and for an illustration of the economics of renewables and electricity storage \cite{prol_economics_2020}. The implementation used here is integrated in a Python-GAMS interface, which enables Python-based data pre- and post-processing, scenario analysis, and visualization \citep{gaete-morales_dieterpy_2021}. For transparency and reproducibility, we provide the model code, all input data, and a manual in public repositories under permissive licenses.\footnote{https://gitlab.com/diw-evu/dieter\_development/dieterpy\_reduced}

We apply the model to a stylized setting loosely parameterized to the German power sector. Annual electricity demand is $520$~TWh, which can be supplied by a mix of two renewable technologies (solar PV and wind power) and two conventional generation technologies (hard coal and open-cycle gas turbines). Assumptions on fixed and variable costs can be found in Table \ref{tab:cost_assumptions} in the supplementary information. We further include a generic electricity storage technology that is parameterized to resemble pumped hydro storage with a round-trip efficiency of 80\%. The energy- and power-related costs of this technology make its application most plausible for medium-duration storage, which suits the cases studied here well. All individual generation units of a technology are modeled as one technology aggregate. This parsimonious model setup allows to focus on the effects of unintended storage cycling for different levels of renewable penetration. Unless stated otherwise, we use a minimum renewable penetration level of 80\%.

\section{An analytical approach to unintended storage cycling}\label{sec:analytical_part}

\subsection{Alternative renewable energy constraint formulations}\label{ssec:res_con_formulation}

Renewable energy constraints in energy models may be generalized as follows:

\begin{equation}\label{eq:general_RES_con}
    base\; element \gtreqless share^{ref} \times reference\; element \pm share^{loss} \times losses
\end{equation}

Varying the base and reference element, four basic constraint formulations are possible. For the sake of argument, we denote them \textit{constraint families}: (1) a minimum renewable share in total demand, (2) a minimum renewable share in total generation; and, conversely, (3) a maximum conventional share in total demand, and (4) a maximum conventional share in total generation. Note that total generation includes storage losses, while total demand does not. Generation from renewables in constraint families (1) and (2) may include both firm and variable renewable energy sources. Formulations (3) and (4) indirectly enforce minimum renewable shares, as the limited generation from conventional technologies has to be complemented by renewables. 

Reflecting different practices in the modeling literature, we investigate three variations for each constraint family as to what degree renewable generators cover storage losses. We denote this characteristic \textit{\ac{SLCR}}. Let $\phi$ be the targeted renewable share, the \ac{SLCR} then may take the following values: (a) storage losses are completely supplied from conventional energy sources (\textit{zero \ac{SLCR}}); (b) storage losses are shared between renewable and conventional energy sources according to the targeted renewable share $\phi$ and $(1-\phi)$, respectively (\textit{proportionate \ac{SLCR}}); or (c) storage losses are completely covered by renewables (\textit{complete \ac{SLCR}}). 

Given current economics, power sector costs generally increase with \ac{VRE} penetration, as the need for providing temporal flexibility rises. Note that we are interested in settings with high \ac{VRE} penetration where the renewable energy constraint is binding, i.e.,~with a penetration beyond the optimal share of an unconstrained model without an explicit renewable energy constraint. Model specifications with zero \ac{SLCR} are the least restrictive ones, as these require the lowest use of \ac{VRE}. In contrast, models with complete \ac{SLCR} are most restrictive, as these require storage losses to be completely covered by \ac{VRE}. Model specifications with proportionate \ac{SLCR} may appear most intuitive and appropriately restrictive. Yet, we show that incomplete \ac{SLCR} levels lead to unintended storage cycling.

In total, the four constraint families and three \ac{SLCR} levels yield twelve different \textit{model specifications} (Table~\ref{tab:res_shares}).

\begin{table}[htbp]
\centering
\ra{1.1}
\caption{Investigated constraint formulations based on Equation~(\ref{eq:general_RES_con}), differentiated by constraint family and \ac{SLCR} level. The right-hand sides of the inequalities define $\Theta$ or $\Omega$ of the optimization problem introduced in Section \ref{ssec:model_definition}.}
\label{tab:res_shares}
\vspace{.1cm}
\small
\begin{tabular*}{\textwidth}{@{\extracolsep{\fill}}lcccccr@{}}\toprule
Model & base     &      & share$^{ref}$ \&  &       & share$^{loss}$ & associated\\
specification & element & sign & reference element         & $\pm$ & storage losses   & constraint \\
\midrule
1a    & $\sum\limits_{s \in \mathcal{R}, t} G_{s,t}$ & $\geq$ & $\phi \sum\limits_t d_t$ & & & (\ref{eq:theta}) \\
1b    & $\sum\limits_{s \in \mathcal{R}, t} G_{s,t}$ & $\geq$ & $\phi \sum\limits_t d_t$ & $+$ & $\phi \sum\limits_{r,t} (G^{in}_{r,t} - G^{out}_{r,t})$ & (\ref{eq:theta}) \\
1c    & $\sum\limits_{s \in \mathcal{R}, t} G_{s,t}$ & $\geq$ & $\phi \sum\limits_t d_t$ & $+$ & $\sum\limits_{r,t} (G^{in}_{r,t} - G^{out}_{r,t})$ & (\ref{eq:theta}) \\
\midrule
2a    & $\sum\limits_{s \in \mathcal{R}, t} G_{s,t}$ & $\geq$ & $\phi \sum\limits_{s,t} G_{s,t}$ & $-$ & $\phi \sum\limits_{r,t} (G^{in}_{r,t} - G^{out}_{r,t})$ & (\ref{eq:theta}) \\
2b    & $\sum\limits_{s \in \mathcal{R}, t} G_{s,t}$ & $\geq$ & $\phi \sum\limits_{s,t} G_{s,t}$ & & & (\ref{eq:theta}) \\
2c    & $\sum\limits_{s \in \mathcal{R}, t} G_{s,t}$ & $\geq$ & $\phi \sum\limits_{s,t} G_{s,t}$ & $+$ & $(1-\phi) \sum\limits_{r,t} (G^{in}_{r,t} - G^{out}_{r,t})$ & (\ref{eq:theta})\\
\midrule
3a    & $\sum\limits_{s \in \mathcal{C}, t} G_{s,t}$ & $\leq$ & $(1 - \phi) \sum\limits_t d_t$ & $+$ & $\sum\limits_{r,t} (G^{in}_{r,t} - G^{out}_{r,t})$ & (\ref{eq:omega}) \\
3b    & $\sum\limits_{s \in \mathcal{C}, t} G_{s,t}$ & $\leq$ & $(1 - \phi) \sum\limits_t d_t$ & $+$ & $(1-\phi) \sum\limits_{r,t} (G^{in}_{r,t} - G^{out}_{r,t})$ & (\ref{eq:omega}) \\
3c    & $\sum\limits_{s \in \mathcal{C}, t} G_{s,t}$ & $\leq$ & $(1 - \phi) \sum\limits_t d_t$ & & & (\ref{eq:omega}) \\
\midrule
4a    & $\sum\limits_{s \in \mathcal{C}, t} G_{s,t}$ & $\leq$ & $(1 - \phi) \sum\limits_{s,t} G_{s,t}$ & $+$ & $\phi \sum\limits_{r,t} (G^{in}_{r,t} - G^{out}_{r,t})$ & (\ref{eq:omega}) \\
4b    & $\sum\limits_{s \in \mathcal{C}, t} G_{s,t}$ & $\leq$ & $(1 - \phi) \sum\limits_{s,t} G_{s,t}$ & & & (\ref{eq:omega}) \\
4c    & $\sum\limits_{s \in \mathcal{C}, t} G_{s,t}$ & $\leq$ & $(1 - \phi) \sum\limits_{s,t} G_{s,t}$ & $-$ & $(1-\phi) \sum\limits_{r,t} (G^{in}_{r,t} - G^{out}_{r,t})$ & (\ref{eq:omega}) \\
\bottomrule
\end{tabular*}
\end{table}

\subsection{Intuition for unintended storage cycling}\label{ssec:intuition}

We first provide some intuition for model specification (1a). Suppose the optimal renewable share of an unconstrained model is below $\phi \sum_t d_t$. Introducing constraint (\ref{eq:theta}) forces the model to increase renewable generation. Further suppose there are hours with renewable curtailment. There are two options to increase $\sum_{s \in \mathcal{R}, t} G_{s,t}$. Either increase renewable generation capacity, which incurs additional investment costs, or transform renewable curtailment into storage losses, which only incurs variable storage costs. The latter is done by increasing $G_{s \in \mathcal{R},t}$ in curtailment hours, and by subsequently charging and discharging storage in these hours. This leads to additional storage losses from unintended storage cycling, which are not accounted for in the renewable constraint formulation of model specification (1a). Suppose demand is 100~MWh in one of these curtailment hours, and storage charging and discharging capacity is 10~MW with a round-trip efficiency of 80\%. Without unintended storage cycling, $G_{s \in \mathcal{R},t}$ in this hour is 100~MWh. Making use of unintended storage cycling, $G_{s \in \mathcal{R},t}$ can increase to 102~MWh, consisting of 10~MWh charged into storage and 92~MWh to satisfy demand. In the same hour, 8~MWh are discharged from storage to serve the load of 100~MWh. In this example, renewable generation increases by 2~MWh compared to a setting without unintended storage cycling. This is done without additional renewable investments, but by transforming 2~MWh of renewable curtailment into storage losses, which allows to meet constraint (\ref{eq:theta}), i.e.,~$\sum_{s \in \mathcal{R}, t} G_{s,t} \geq \phi \sum_t d_t$, at lower costs. Note that the additional 2~MWh of renewable generation contribute to meeting the renewable energy target (Equation~(\ref{eq:theta})), but do not serve electricity demand (Equation~(\ref{eq:energy_balance})), as the energy is lost in the storage conversion process.

Under model specification (1b), a similar reasoning applies. The only difference is that storage losses are partially accounted for in the renewable constraint with $\phi \sum_{r,t} (G^{in}_{r,t} - G^{out}_{r,t})$. Only the remaining fraction of storage losses, $(1-\phi) \sum_{r,t} (G^{in}_{r,t} - G^{out}_{r,t})$, can be used to transform curtailment into storage losses via unintended storage cycling. Suppose the targeted renewable share is $\phi = 80\%$ in the setting described above. Unintended storage cycling still leads to a renewable generation increase of 2~MWh in the exemplary hour, which is transformed into storage losses. Yet, 80\% of these losses are accounted for in the renewable constraint. Thus, only 20\% of the increase in renewable generation, i.e.,~0.4~MWh, help to relax the renewable constraint. To achieve a similar effect as under model (1a), more unintended storage cycling would be required. 

Under model specification (1c), storage losses are completely covered by the renewable energy constraint. Unintended storage cycling would only increase total variable storage costs, without providing any benefit in terms of meeting the renewable constraint. Accordingly, renewable generation can only be increased via additional renewable capacity investments.

A similar reasoning applies to the other renewable share constraints listed in Table~\ref{tab:res_shares}. For models using constraint family (2), the only difference is that the fraction of storage losses covered by renewables is slightly less intuitive than in the example described above, as storage losses are already included in overall generation $\sum_{s,t} G_{s,t}$. The effects in constraint families (3) and (4) are similar to the ones in (1) and (2), respectively. 

\subsection{Long-term equilibrium conditions for storage}\label{ssec:lt_equilibrium}

We first consider an unconstrained long-term equilibrium of the model defined in Section \ref{ssec:lt_equilibrium} without binding renewable energy targets, i.e.,~discarding constraints (\ref{eq:omega}) and (\ref{eq:theta}). Joskow in~\cite{joskow_comparing_2011} stresses that both cost and value of a technology matter in such a setting. The \ac{LCOS} of a storage unit~$r$ is the average cost of each unit energy discharged \cite{pawel_cost_2014}. It includes investment and variable costs as well as the cost of the electricity used for charging. As it relates to the storage output, the \ac{LCOS} also consider the value of conversion losses. 

\begin{equation}\label{eq:lcos}
    LCOS_r = \frac{\sum_\circ i_r^\circ C_r^\circ + \sum_{t,*} o_r^* G_{r,t}^* + \sum_t \lambda_t G_{r,t}^{in}}{\sum_t G^{out}_{r,t}}
\end{equation}
\vspace{0.1cm}

When interpreting the \ac{LCOS} metrics, the temporal heterogeneity of electricity as a good has to be considered. The value of electricity is time-variant due to a number of reasons \cite{joskow_comparing_2011,hirth_market_2013,ueckerdt_system_2013}. Both the electricity demand and the output of many generation technologies are fluctuating in nature. Especially the value of \ac{VRE} depends on the time when they generate, i.e.,~their variable availability profile \cite{prol_economics_2020}, and the location of generation in the geographical context. Consequently, the electricity price fluctuates over time and space, too. Transmission and distribution grid constraints might further affect the locational value of \ac{VRE}. 

The \ac{MV} accommodates these variations in electricity value \cite{joskow_comparing_2011}. The \ac{MV} of a storage technology~$r$ refers to the average value of each unit of energy supplied to the system. Storage operators generate revenue by dispatching stored energy at market prices $\lambda_t$. They exploit arbitrage opportunities, ideally dispatching during periods with high prices, while charging during low-price periods \cite{brown_decreasing_2021}. 

\begin{equation}\label{eq:mv_r}
    MV_r = \frac{\sum_t \lambda_t G_{r,t}^{out}}{\sum_t G^{out}_{r,t}}
\end{equation}
\vspace{0.1cm}

Storage achieves economic efficiency if \ac{LCOS} equals its marginal economic value, and competitiveness if \ac{LCOS} is less or equal to its \ac{MV}. For perfect and complete markets, the marginal economic value and the \ac{MV} coincide \cite{ueckerdt_system_2013}. In the long-term equilibrium of an unconstrained optimum, it follows that \ac{LCOS} equal storage \ac{MV}, i.e.,~storage operators generate zero profit as costs perfectly match revenue (zero-profit condition) \cite{brown_decreasing_2021}. Using the Lagrangian function and \ac{KKT} conditions (Section~\ref{sec:lagrange}), Section \ref{ssec:derivations_unconstrained_opt} proves this optimality condition. 

\begin{equation}
    LCOS_r = MV_r
\end{equation}

Imposing a binding renewable energy target, i.e.,~including constraints (\ref{eq:omega}) and (\ref{eq:theta}), may alter the zero-profit condition of storage in a long-term equilibrium. For clarity, we define a new metric, the \ac{NSL} of a storage unit~$r$. It indicates storage conversion losses related to each unit of storage output. 

\begin{equation}\label{eq:nsl}
    NSL_r = \frac{\sum_t \big( G_{r,t}^{in} - G_{r,t}^{out}\big)}{\sum_t G^{out}_{r,t}}
\end{equation}
\vspace{0.1cm}

Table \ref{tab:zero_profit_r} shows that a binding renewable energy target may affect the composition of storage costs and revenue streams of all investigated model specifications. Exemplary proofs of those conditions can be found in Section~\ref{ssec:derivations_constrained_opt}.

Within each constraint family, the \ac{LCOS} increase by a certain margin, the more storage losses are covered by renewables. This margin is a multiple of the factor $\mu_{\theta/\omega} NSL_r$, which represents the storage loss per unit storage output, valued at the marginal system costs of an additional unit of renewable generation. For a binding renewable target, renewable energy becomes a scarce good with additional inherent value. When renewable energy is lost in storage conversion processes, its inherent value for achieving the renewable constraint is annihilated. Restoring the lost energy incurs additional costs. We thus interpret this factor as storage integration costs that arise in a system with a binding renewable energy constraint. If the constraint is not binding, the dual variable of the minimum renewable constraint $\mu_{\theta/\omega}$ is zero, and the zero-profit condition of storage remains unchanged compared to the unconstrained optimum without renewable policy targets.

\begin{table}[htbp]
\centering
\ra{1.1}
\caption{Storage operators' zero-profit conditions.}
\label{tab:zero_profit_r}
\small
\begin{tabular}{lllr}
\toprule
model & costs & & revenue \\
\midrule
1a  & $LCOS_r$ & $=$ & $MV_r$ \\
1b  & $LCOS_r + \phi \mu_{\theta} NSL_r $ & $=$ & $MV_r$\\
1c  &$LCOS_r + \mu_{\theta} NSL_r $ & $=$ & $MV_r$ \\
\midrule
2a, 4a  & $LCOS_r -\phi \mu_{\theta} NSL_r $ & $=$ & $ MV_r $\\
2b, 4b  & $LCOS_r $ & $=$ & $ MV_r $ \\
2c, 4c  & $LCOS_r + (1 - \phi) \mu_{\theta} NSL_r $ & $=$ & $ MV_r$ \\
\midrule
3a  & $LCOS_r - \mu_{\omega} NSL_r$ & $=$ & $ MV_r$ \\
3b  & $LCOS_r - (1 - \phi) \mu_{\omega} NSL_r$ & $=$ & $ MV_r $ \\
3c  & $LCOS_r $ & $=$ & $ MV_r $ \\
\bottomrule
\end{tabular}
\end{table}

Models with complete \ac{SLCR} include all storage losses arising in the system in the formulation of the renewable energy constraint. The zero-profit condition of storage operators properly accounts for all storage loss costs related to the binding renewable target. The economic trade-off between storage use and renewable curtailment is properly specified such that unintended storage cycling is prevented. 

In contrast, models with zero \ac{SLCR} neglect storage losses in the minimum renewable constraint. They can be completely covered by conventional generators. Consequently, the zero-profit condition of storage operators does not include any storage loss costs related to the binding renewable target. Within in the same constraint family, the zero-profit condition is reduced by the factor $\mu_{\theta/\omega} NSL_r$ compared to models with complete \ac{SLCR}. Similarly, models with proportionate \ac{SLCR} only partially require renewables to cover storage losses. The system costs of storage losses are thus only partially imposed on storage operators in the long-term equilibrium. Compared to models with complete \ac{SLCR} within in the same constraint family, the zero-profit condition is reduced by the factor $(1-\phi) \mu_{\theta/\omega} NSL_r$, which corresponds to the share of storage loss costs that is not accounted for in the renewable constraint. In other words, models based on zero and proportionate \ac{SLCR} do not fully account for the underlying \ac{LCOS} related to the binding renewable target. This causes excessive storage use and too little \ac{VRE} surplus curtailment in the long-term equilibrium.

\section{An illustration of different types of unintended storage cycling}\label{sec:usc_types}

Unintended storage cycling can be observed in case of simultaneous charging and discharging of storage. This involves energy that is cycled through the storage within the same period, which we refer to as \ac{SPC}. Further, there is also an inter-temporal instance of the artifact: \ac{APC}. It refers to a situation in which unintendedly discharged energy has been charged prior to the period of unintended discharge. Both \ac{SPC} and \ac{APC} are excessive and constitute unintended storage use.

\begin{figure}[htbp]
\centering
\subfloat[Type 1: Storage charging equals discharging.\label{fig:sto_cyc_type1}]
    {{\includegraphics[width=0.47\textwidth]{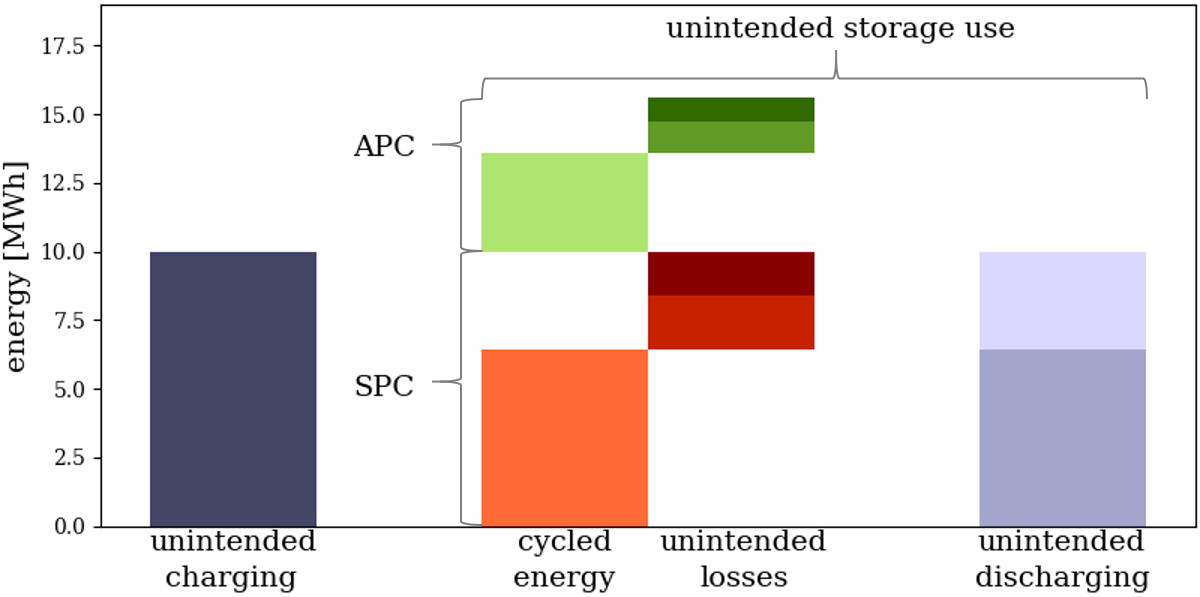}}}%
\qquad
\subfloat[Type 2: Storage discharging exceeds charging.\label{fig:sto_cyc_type2}]
    {{\includegraphics[width=0.47\textwidth]{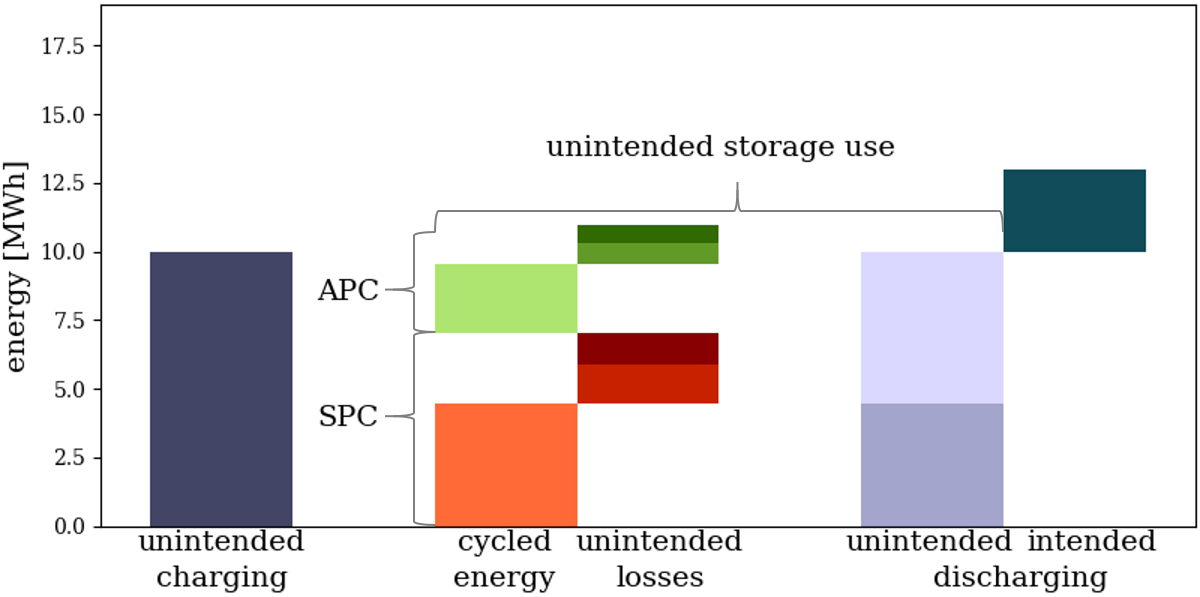}}}\\%
\subfloat[Type 3: Storage charging exceeds discharging; discharging exceeds same-period unintendedly cycled energy.\label{fig:sto_cyc_type3}]
    {{\includegraphics[width=0.47\textwidth]{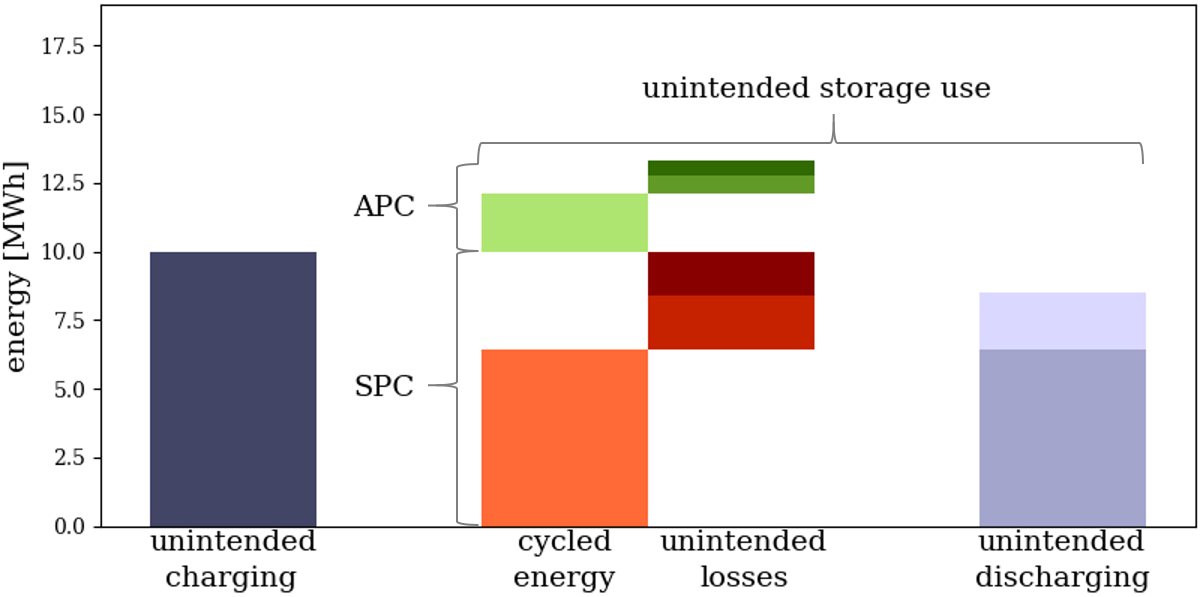}}}
\qquad
\subfloat[Type 4: Storage charging exceeds discharging; but not the full cycling potential is exploited.\label{fig:sto_cyc_type4}]
    {{\includegraphics[width=0.47\textwidth]{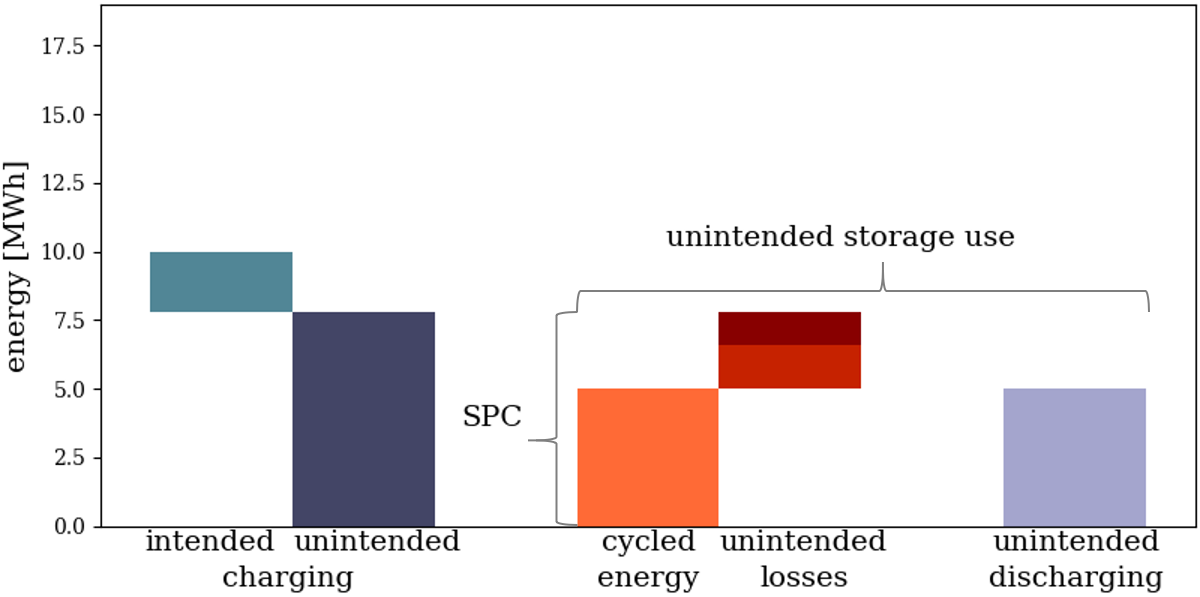}}}\\
\subfloat{\includegraphics[width=\textwidth]{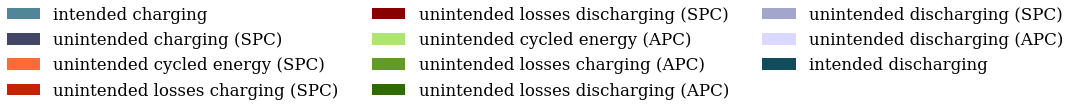}}
\caption{Four types of unintended storage cycling.}%
\label{fig:sto_cyc_types}%
\end{figure}

There are four conceivable types of unintended storage cycling. They differ in terms of the ratio of storage charging to discharging within the same period. Further, they vary in terms of the temporality of unintendedly cycled energy, which may be cycled either within the period (\ac{SPC} only), or within the same and across periods (\ac{SPC} and \ac{APC}). Figure \ref{fig:sto_cyc_types} illustrates these types with stylized numbers. For the sake of illustration, they are based on an assumed low efficiency for charging and discharging of 80\%, respectively.

The first type occurs in a situation in which storage charging and discharging within the same period are equal (Figure \ref{fig:sto_cyc_type1}). Due to conversion losses, out of 10~MWh charged energy (left bar) only 6.4~MWh can be discharged (dark area, right bar). This energy amount represents the unintendedly cycled energy within the same period (orange area, middle bar). Corresponding losses when charging and discharging amount to 2.0 and 1.6~MWh (light and dark red areas, middle bar), respectively. We define the total energy required for cycling these 10~MWh within the same hour as \ac{SPC}. It comprises the unintendedly cycled energy (6.4~MWh) plus conversion losses (3.6~MWh in total), totaling 10~MWh. To achieve a total discharge of 10~MWh in this period, it takes another 3.6~MWh (lighter area, right bar) of energy that already had to be stored during a previous period. This additional unintendedly cycled energy of 3.6~MWh (light green, middle bar) is cycled across periods. Again, it entails storage conversion losses when charging (1.125~MWh, medium green, middle bar) and discharging (0.9~MWh, dark green, middle bar). We define the energy for unintended cycling across periods (3.6~MWh plus 2.025~MWh losses), totaling 5.625~MWh, as \ac{APC}. The total energy required for equal storage charging and discharging of 10~MWh within the same period is 25.625~MWh. It contains the total charging, which is the sum of \ac{SPC} and \ac{APC} (15.625~MWh); plus the total unintended discharging (10~MWh). The net effect of such storage use on the system's energy balance in the respective hour is zero. Thus, the entire 25.625~MWh are considered as unintended storage use. We then define unintended storage losses as the energy removed from the system through \ac{SPC} and \ac{APC} losses. Essentially, unintended storage cycling transforms \ac{VRE} surplus into unintended storage losses and circumvents renewable curtailment. In our running example, the unintended storage losses is 5.625~MWh.

The second type of unintended storage cycling occurs when storage discharge exceeds the same-period charging, i.e.,~effective discharging (Figure \ref{fig:sto_cyc_type2}). Both \ac{SPC} and \ac{APC} occur in this case, similar to the first type. However, the difference between the same-period charging and discharging (dark area, right bar) exerts a net effect on the system's energy balance in this period. This is intended storage use that helps to serve demand, and needs to be distinguished from storage use solely arising to generate additional losses to remove excess electricity from the system.

The third and fourth types of unintended storage cycling occur when more energy is charged than discharged within the same period. In the third type, total discharged energy consists of both \ac{SPC} and \ac{APC}, with less use of \ac{APC} compared to the first type (Figure \ref{fig:sto_cyc_type3}). In the fourth type, effective charging occurs. There is no \ac{APC}, and the unintended cycling potential is not necessarily fully exhausted (Figure \ref{fig:sto_cyc_type4}). In the depicted case, the \ac{SPC} is smaller than the charged energy, leaving room for some level of intended charging.

\section{Numerical results for unintended storage cycling}\label{sec:numerical_results}

\subsection{Occurrence and effects on total system costs}\label{ssec:results_intro}

\begin{figure}[htbp]%
\centering
\subfloat[Number of hours in which unintended storage cycling occurs.\label{fig:sto_cyc_h}]
    {{\includegraphics[width=0.47\textwidth]{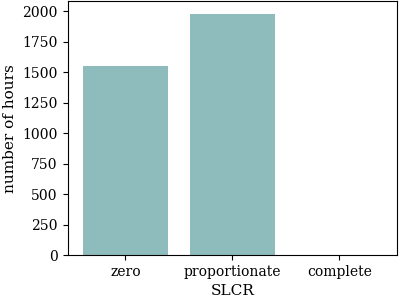}}}
\qquad
\subfloat[Total systems costs.\label{fig:tc}]
    {{\includegraphics[width=0.47\textwidth]{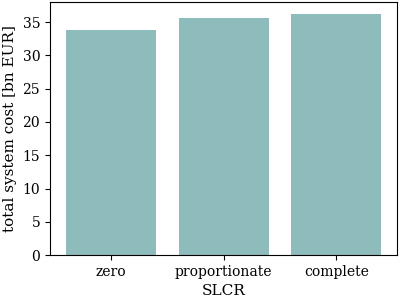}}}
\caption{Number of hours in which unintended storage cycling occurs and total system costs per \ac{SLCR} level at a minimum renewable share of 80\%. These graphs are similar for all constraint families, with only marginal differences.}%
\label{fig:medical_report}%
\end{figure}

We test 12 different model specifications with alternative minimum renewable share constraint formulations as detailed in Table~\ref{tab:res_shares}, using the stylized numerical model introduced in Section~\ref{ssec:numerical_model_defintion}. We find that unintended storage cycling is a frequently occurring modeling artifact. It arises when the \ac{SLCR} level of the imposed renewable share constraint is incomplete (i.e.,~constraints with suffix a and b in Table~\ref{tab:res_shares}), regardless of the constraint family (1, 2, 3, or 4). This is most pronounced for models with proportionate \ac{SLCR}. Only models with complete \ac{SLCR} (constraints c) prevent the model artifact (Figure \ref{fig:sto_cyc_h}), i.e., requiring renewable generation to completely cover storage losses averts unintended storage cycling. We further find the fourth type of unintended storage cycling to be most common. Most model outcomes vary across \ac{SLCR} levels (i.e.,~constraints with suffix a, b, or c), but coincide across constraint families. For the sake of illustration, we thus present our findings differentiated by \ac{SLCR} level, unless stated otherwise. 

Confirming the intuition laid out in the previous sections, the artifact leads to lower-cost solutions (Figure~\ref{fig:tc}). This essentially happens as the renewable constraint is partly relaxed in the sense that storage losses also contribute to meeting the renewable target. Compared to the setting with complete \ac{SLCR}, some renewable curtailment is avoided and transformed into additional \ac{VRE} generation, and subsequently, into storage losses. However, this energy is not used to serve final demand - which is clearly an unintended effect of the renewable constraint formulation. 

\subsection{Effects on optimal dispatch and investment decisions}\label{ssec:results_dispatch_investment}

\begin{figure}[htbp]
\centering
\subfloat[Model specification (1a) with zero \ac{SLCR}.\label{fig:profile_1a}]
    {{\includegraphics[width=\textwidth]{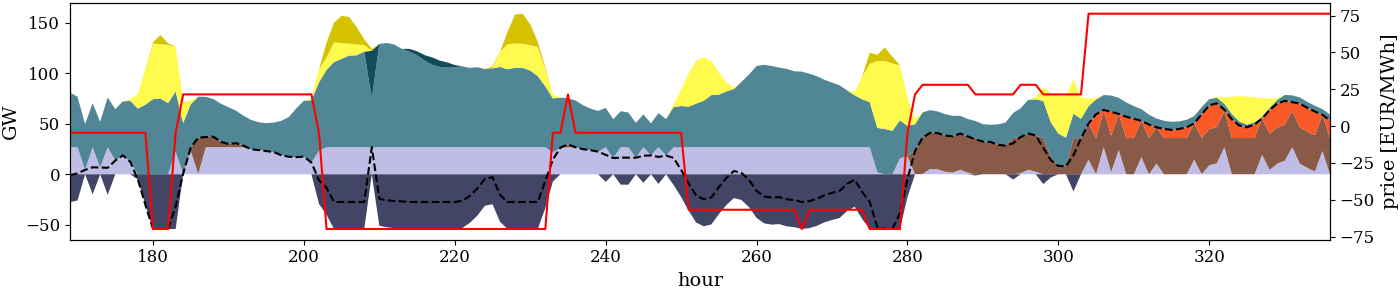}}}\\
\subfloat[Model specification (1b) with proportionate \ac{SLCR}.\label{fig:profile_1b}]
    {{\includegraphics[width=\textwidth]{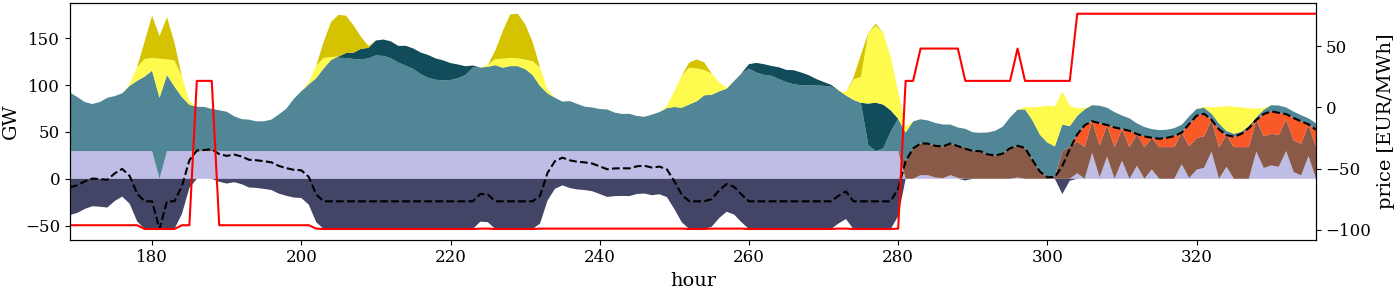}}}\\
\subfloat[Model specification (1c) with complete \ac{SLCR}.\label{fig:profile_1c}]
    {{\includegraphics[width=\textwidth]{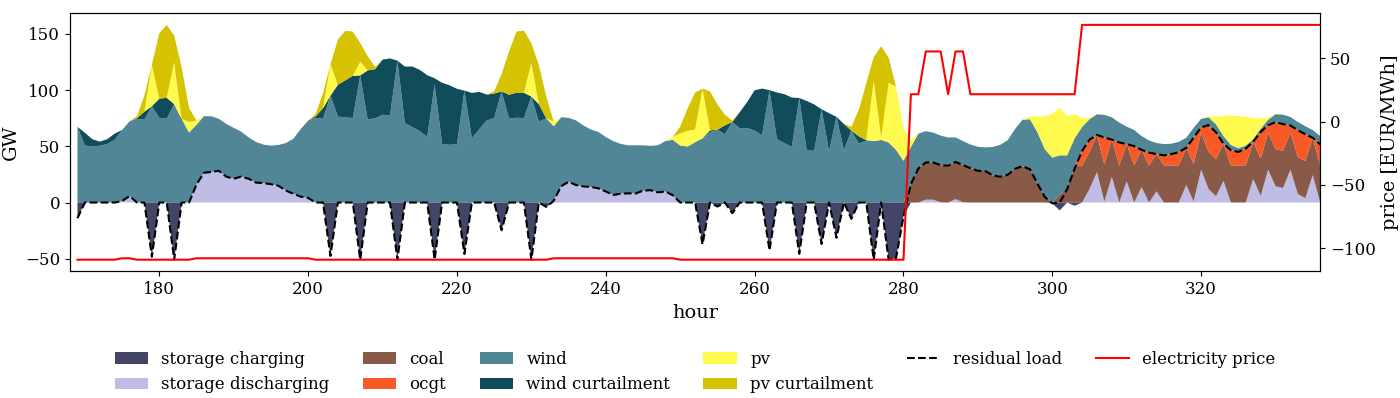}}}\\
\caption{Generation and price profiles for the second week of the target year at a renewable target of 80\% in total demand. The positive part of the ordinate relates to storage discharging, its negative part to charging.}%
\label{fig:gen_profiles}%
\end{figure}

Figure \ref{fig:gen_profiles} illustrates generation, curtailment, and storage use for an exemplary week for model specifications imposing a minimum renewable share in demand with zero \ac{SLCR} (1a), proportionate \ac{SLCR} (1b), and complete \ac{SLCR} (1c). Respective illustrations for the other three constraint families (2, 3, 4) are very similar with only marginal deviations. In the models with zero and proportionate \ac{SLCR}, unintended storage cycling takes place, e.g.,~between hours 200 and 230, but not in the model with complete \ac{SLCR}. The artifact's impact is not limited to the simultaneity of charging and discharging. The hourly storage usage patterns also differ substantially.

Unintended storage cycling also affects the generation profiles of renewable and conventional generators, as well as residual load. The latter is demand for electricity in a given time period, e.g.,~one hour, net of \ac{VRE} generation potential during this period. A positive residual load refers to situations in which demand exceeds the \ac{VRE} generation potential, and requires the use of dispatchable generation technologies. In contrast, the residual load is negative if the \ac{VRE} generation potential exceeds demand. Residual load in the exemplary week is lower during periods of renewable curtailment in models with zero or proportionate \ac{SLCR} compared to models with complete \ac{SLCR}. This is a consequence of lower curtailment of renewable generation, notably during the first days of the week, which is transformed into additional storage losses related to unintended storage cycling. During the last days of the week, renewable energy is scarce as a result of prevailing weather conditions, and residual load has to be met by dispatchable power plants. Generation from conventional technologies in models with zero or proportionate \ac{SLCR} slightly exceed their complete \ac{SLCR} equivalent. This is caused by lower \ac{VRE} capacity in the models with incomplete \ac{SLCR}.

\begin{figure}[htbp]%
\centering
\subfloat[Installed generation capacity including storage charging and discharging.\label{fig:cap}]
    {{\includegraphics[width=0.47\textwidth]{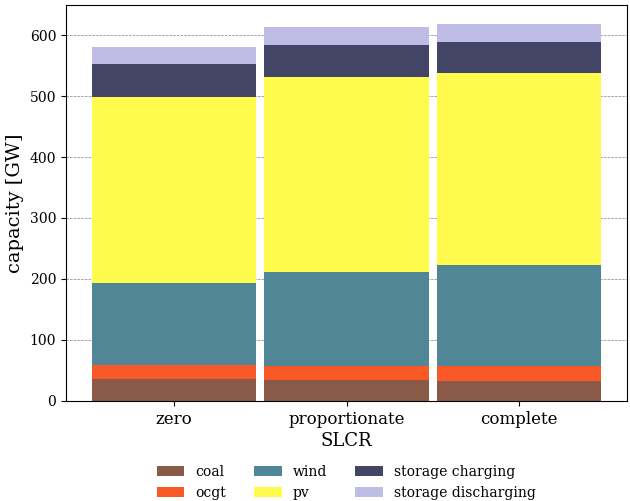}}}
\qquad
\subfloat[Installed storage energy capacity.\label{fig:cap_sto_e}]
    {{\includegraphics[width=0.47\textwidth]{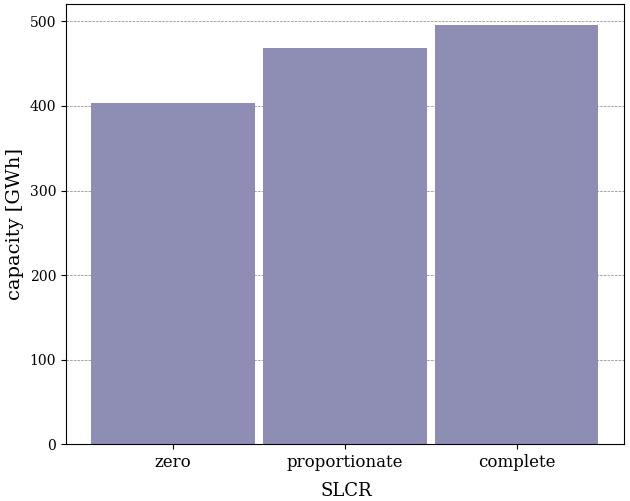}}}\\
\subfloat[Generation including storage charging (negative part of ordinate) and discharging.\label{fig:gen}]
    {{\includegraphics[width=0.47\textwidth]{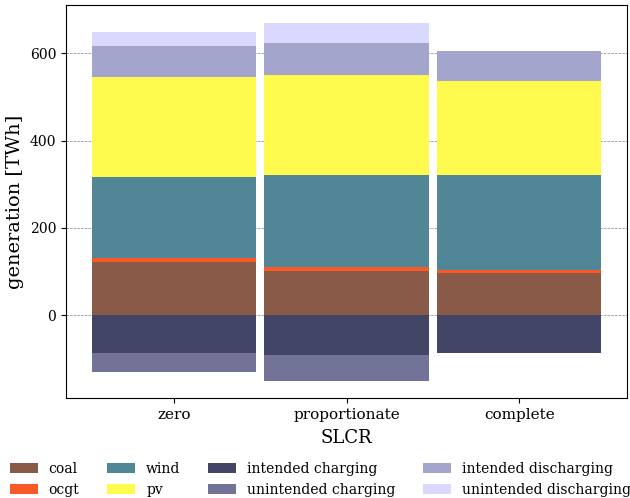}}}
\qquad
\subfloat[Storage losses and curtailment.\label{fig:sto_loss_cu}]
    {{\includegraphics[width=0.47\textwidth]{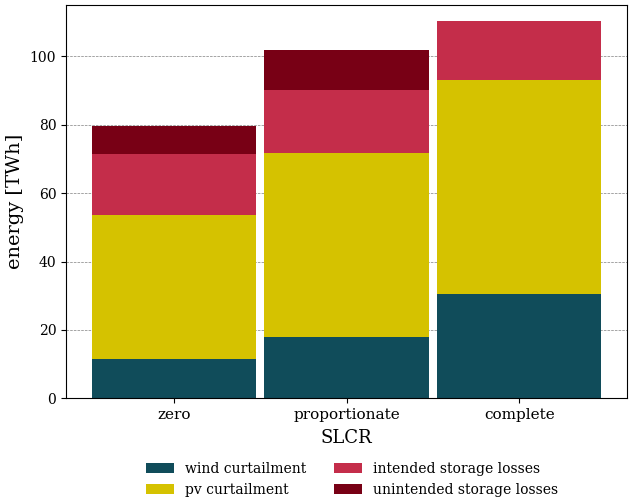}}}
\caption{Installed capacity, annual generation, curtailment, and storage losses per technology at a renewable share of 80\%.}
\label{fig:gen_cap}%
\end{figure}

Unintended storage cycling also distorts optimal capacity and annual dispatch decisions. The effects are largely similar in models using the same \ac{SLCR} level across all constraint families.\footnote{While investment decisions coincide across all constraint families (Figure~\ref{fig:cap_annual}), constraint family~(3) models use a bit less wind power and more \ac{PV} (Figure~\ref{fig:gen_annual}).} In models with incomplete \ac{SLCR}, \ac{VRE} generation capacity (Figure~\ref{fig:cap}) and storage energy capacity (Figure~\ref{fig:cap_sto_e}) decrease, while the dispatch of conventional generators as well as storage use increase (Figure~\ref{fig:gen}). Renewable curtailment declines, and is partly converted into unintended storage losses (Figure~\ref{fig:sto_loss_cu}).

If unintended storage cycling occurs, the increase in total storage charging and discharging is related to unintended storage use (Figure~\ref{fig:gen}), generating unintended storage losses (Figure~\ref{fig:sto_loss_cu}). This effect is most pronounced in models with proportionate \ac{SLCR}, as only the fraction of the additional storage losses helps to achieve the renewable target that is not covered by the renewable energy constraint (compare Section~\ref{ssec:intuition}). About 96\% of the unintended storage use relates to \ac{SPC}, while \ac{APC} accounts for the remaining 4\%.

Renewable curtailment decreases due to the conversion of renewable curtailment into additional \ac{VRE} generation, and subsequently, unintended storage losses. This substitution is most effective in  models with zero \ac{SLCR}. 
Because of the additional renewable generation that is not curtailed, but removed via unintended storage cycling, much less \ac{VRE} capacity is required in models with incomplete \ac{SLCR} to meet the renewable energy constraint (Figure~\ref{fig:cap}). This is most pronounced for wind power, which is the most expensive renewable energy source under the parameterization used here. \ac{PV} capacity slightly increases, while optimal wind power deployment disproportionately decreases compared to the case with complete \ac{SLCR}. The same holds true for generation from both technologies (Figure~\ref{fig:gen}). Renewable curtailment decreases further as a consequence of the capacity effect. Accordingly, renewable surplus energy declines, such that the need for storing excess energy over a longer period of time decreases, which yields a lower optimal storage energy capacity (Figure~\ref{fig:cap_sto_e}).

Renewable energy that it is removed from the system in the form of unintended storage losses contributes to the renewable energy constraint. However, it does not provide useful energy for serving demand. To meet the system's energy balance, generation from coal plants thus increases during hours in which, due to the \ac{VRE} capacity effect, the \ac{VRE} generation potential decreases compared to a setting without unintended storage cycling. This is most pronounced for models with zero \ac{SLCR}.
Because of the increase in generation from coal plants, carbon dioxide emissions grow by 25\% or 6\% in models with zero or proportionate \ac{SLCR}, respectively. 

Aside from effects directly related to unintended storage cycling, the developments illustrated in Figure~\ref{fig:gen_cap} are also driven by a lower ambition level of the renewable energy constraints of models with zero or proportionate \ac{SLCR}. Since storage losses can, at least partly, be covered by conventional generators here, these models require lower renewable generation than a specification with complete \ac{SLCR} by definition. Section~\ref{ssec:factor_separation} disentangles these the impact of both factors. The factor separation raises complementary insights, but also introduces new complexities. For practical model applications, only the combined effects of varying renewable ambition levels and unintended storage cycling are relevant, which are shown in Figure~\ref{fig:gen_cap}.

\subsection{System effects of storage illustrated with residual load duration curves}

\begin{figure}[htbp]%
\centering
\subfloat[Model specification (1c) with complete \ac{SLCR}.\label{fig:rldc_1c}]
    {\includegraphics[width=\textwidth,keepaspectratio]{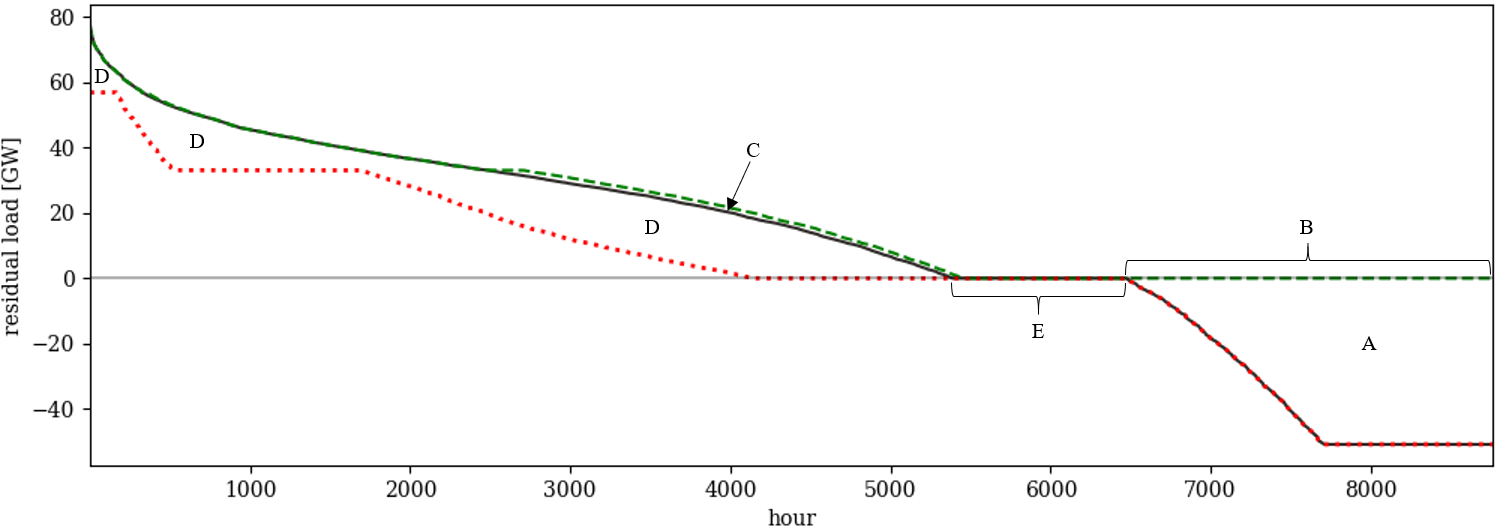}}\\
\subfloat[Model specification (1b) with proportionate \ac{SLCR}.\label{fig:rldc_1b}]
    {{\includegraphics[width=\textwidth]{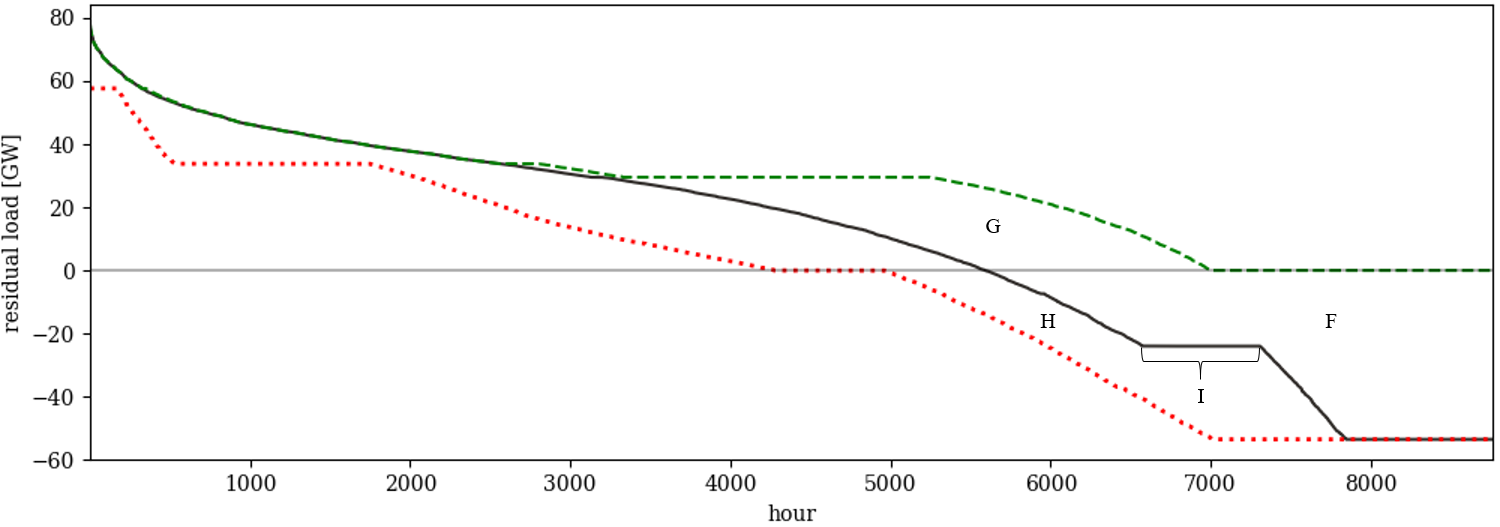}}}\\
\subfloat{\includegraphics[width=0.45\textwidth]{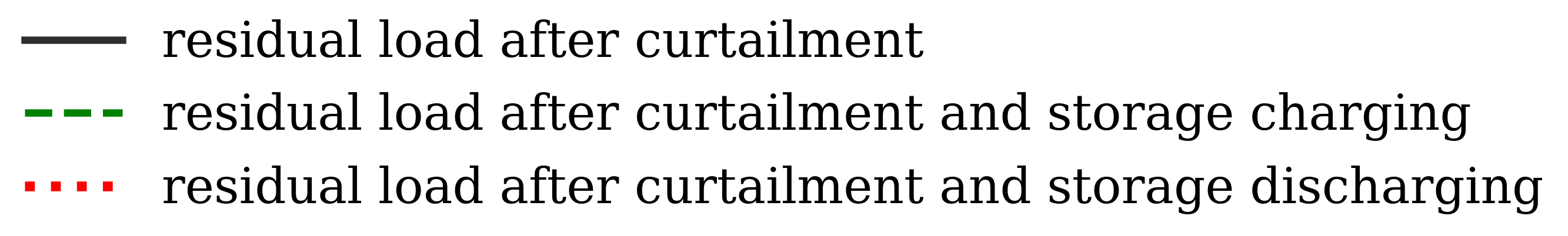}}
\caption{Different configurations of the \ac{RLDC} at a targeted renewable share of 80\%.}%
\label{fig:rldc}%
\end{figure}

\begin{figure}[htbp]
\centering
\noindent\includegraphics[width=\linewidth,height=\textheight, keepaspectratio]{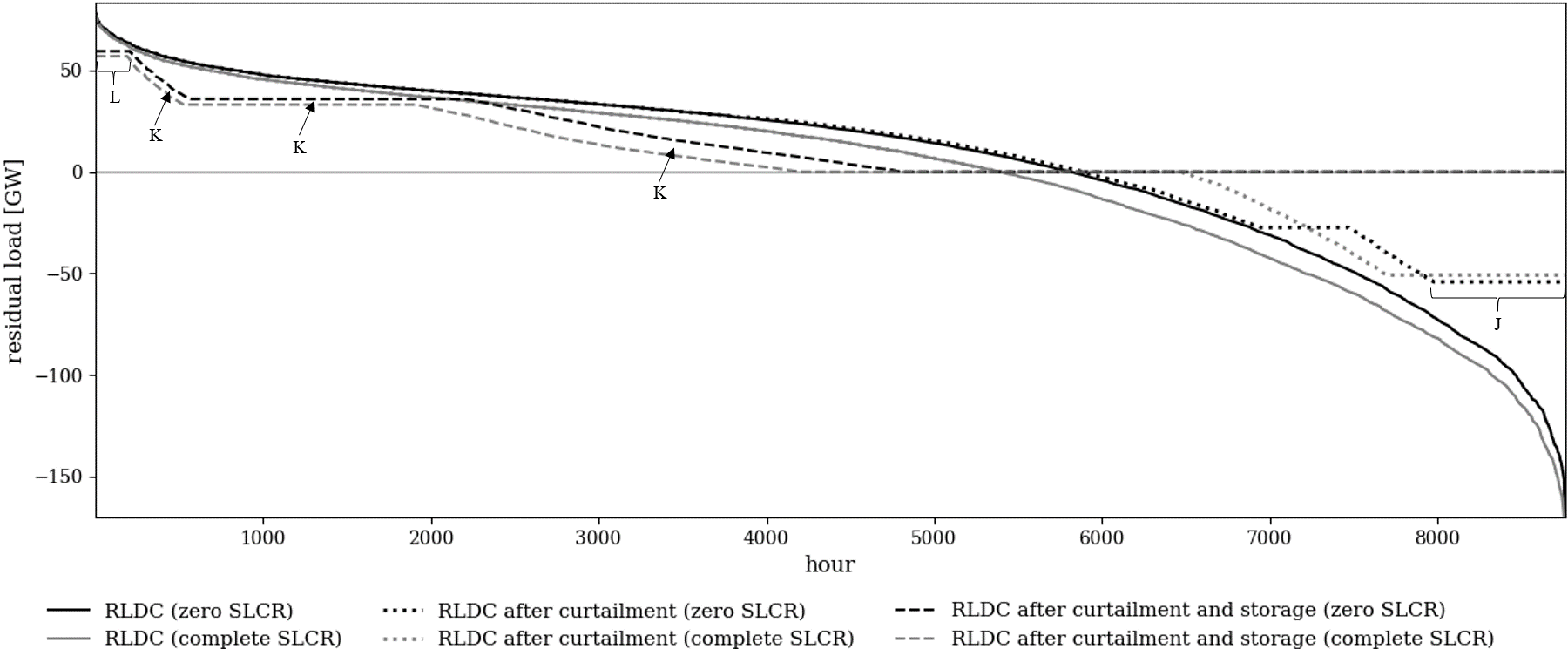}
\caption{Different configurations of the \ac{RLDC} for the model specification (1a) with zero \ac{SLCR} in black and the specification (1c) with complete \ac{SLCR} in gray at a targeted renewable share of 80\%.}
\label{fig:rldc_1ac}
\end{figure}

To illustrate system effects of electricity storage and \ac{VRE}, a \ac{RLDC} can be used \cite{zerrahn2018,schill_electricity_2020,ueckerdt_representing_2015}. A standard \ac{RLDC} sorts all hourly residual load values of a full year in descending order (solid lines in Figure~\ref{fig:rldc_1ac}). With increasing \ac{VRE} penetration, the \ac{RLDC} shifts downwards especially on the right-hand side \cite{prol_economics_2020}. Augmented versions of residual load duration curves can be calculated that also consider optimal curtailment and electricity storage use (Figures~\ref{fig:rldc} and \ref{fig:rldc_1ac}). 

In a case without unintended storage cycling, in the optimum storage takes up a certain amount of renewable surplus generation (area A in Figure~\ref{fig:rldc_1c}, leading to the plateau B) as well as some generation from dispatchable power plants with low variable costs (area C). Storage shifts this energy to periods with positive residual load and \ac{VRE} shortage (area D), displacing conventional generation. Optimal renewable curtailment yields a number of situations with zero residual load in which storage is not being used (plateau E of the \ac{RLDC} after curtailment in Figure~\ref{fig:rldc_1c}).

Accordingly, \ac{RLDC}s illustrate the trade-off between optimal storage sizing and \ac{VRE} curtailment. As storage investments incur costs, it is not optimal to fully take up renewable surplus generation, but to also allow some \ac{VRE} curtailment. In general, this lowers the need for both storage charging and energy capacity \cite{zerrahn2018,schill_residual_2014}.

Unintended storage cycling distorts these optimal curtailment and storage use patterns (Figure~\ref{fig:rldc_1b}\footnote{The Figure shows the case with proportionate \ac{SLCR}. The \ac{RLDC} of a model specification with zero \ac{SLCR} is similarly affected and omitted for the sake of clarity. The \ac{RLDC}s of constraint family~(3) models deviate only marginally.}). A larger part of \ac{VRE} surplus generation is taken up by storage and not curtailed (compare area in F in Figure~\ref{fig:rldc_1b} to area A in Figure~\ref{fig:rldc_1c}). Unintended storage cycling results in energy being charged (area G in Figure~\ref{fig:rldc_1b}) and discharged again in the same period (area H). In contrast to the case without unintended storage cycling, there are no hours left in which \ac{VRE} surplus is curtailed without storage being charged, which would result in a plateau of the residual load at the zero line. Instead, there are hours in which maximum possible unintended storage losses are generated (plateau I). Here, both storage charging and discharging are at maximum capacity, while the remaining \ac{VRE} surplus is curtailed. Note that same-period storage cycling is limited not only by the installed storage charging capacity, but also by the discharging capacity. In our parameterization, optimal storage discharging capacity is lower than the optimal charging capacity. In hours to the left and right of those in plateau I, unintended storage cycling may also occur. Yet, not the full cycling potential is exploited here. On the right-hand side of plateau I, storage charging capacity is never at the maximum, while storage discharging capacity is never at the maximum on the left-hand side of plateau I.

Optimal dispatch and investment decisions of all technologies are affected by unintended storage cycling, which is most pronounced in settings with zero \ac{SLCR}. Here, optimal storage charging capacity slightly increases to expand the unintended storage cycling potential (vertical distance between dashed lines in plateau J in Figure~\ref{fig:rldc_1ac}). Further, the \ac{VRE} capacity is lower than in models with complete \ac{SLCR}, yielding an upward shift of the \ac{RLDC} (area between solid lines in Figure~\ref{fig:rldc_1ac}\footnote{While less pronounced, similar effects also occur in settings with proportionate \ac{SLCR}. Yet, for better illustration, we contrast \ac{RLDC}s of models with zero and complete \ac{SLCR}.}). While the shifted \ac{VRE} surplus energy from the negative part of the \ac{RLDC} increases, the additionally charged energy is not used to serve demand, but partly removed from the system via unintended storage cycling. As a consequence, conventional generation increases (area K between the dashed lines in Figure~\ref{fig:rldc_1ac}).
This goes along with additional capacity entry of coal in models with zero \ac{SLCR}, which helps to serve residual load during peak periods (vertical distance between dashed lines in plateau L in Figure~\ref{fig:rldc_1ac}). 

\subsection{Effects on market clearing prices}\label{ssec:results_prices}

The merit order ranks available generation technologies in ascending order of variable generation costs. In our cost-minimizing setting, technologies with lowest variable costs are dispatched first to meet demand. In the absence of electricity storage, the technology that supplies the last unit of demand --- the marginal technology --- sets the uniform price for the power sector at the intersection of demand and supply. The dual variable $\lambda_t$ of the energy balance, Equation~(\ref{eq:energy_balance}), indicates the resulting price, which depends on the optimality condition of the marginal technology. It can be derived from the first-order condition of the underlying optimization model that applies to the generation variable of the marginal technology \cite[Ch.4]{biggar_economics_2014}. The price constitutes the sum of variable costs and the dual variables of constraints that may be binding for the marginal technology. 


If storage is deployed, the first-order condition of storage discharging includes variable costs, the dual variable of capacity- and energy-related constraints, and the derivative of the renewable energy constraint (Equations~(\ref{eq:kkt_G_r_out_theta}) and~(\ref{eq:kkt_G_r_out_omega})). While most of these summands coincide for all model specifications, the derivative of the renewable constraint may vary over both constraint families and \ac{SLCR} levels (Equations~(\ref{eq:opt_cond_sto_out_1a}) to (\ref{eq:opt_cond_sto_out_4c})). 

Similarly, the first-order condition of storage charging determines the price at which storage charges (Equations~(\ref{eq:kkt_G_r_in_theta}) and~(\ref{eq:kkt_G_r_in_omega})). This is the price that results at the intersection of demand and supply. It may also change across both constraint families and \ac{SLCR} levels due to a variation in the derivation of the renewable energy constraint with respect to storage charging (Equations~(\ref{eq:opt_cond_sto_in_1a}) to (\ref{eq:opt_cond_sto_in_4c})). 

\begin{figure}[htbp]
\centering
\noindent\includegraphics[width=0.65\linewidth,keepaspectratio]{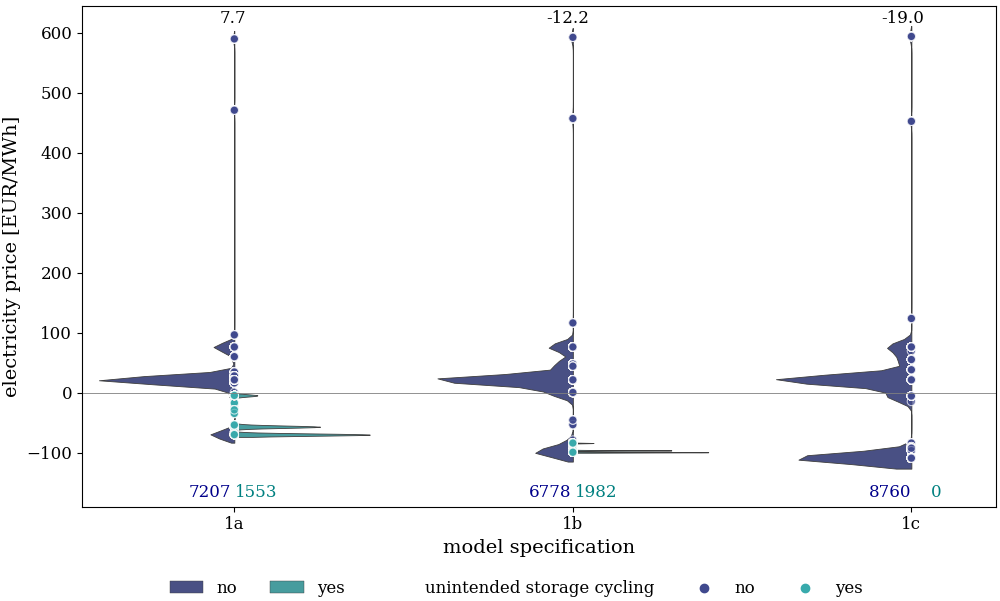}
\caption{Distribution of prices across different \ac{SLCR} levels for a minimum renewable share in demand. The left side of each violin in dark blue shows the distribution of prices during hours without simultaneous storage charging and discharging. The right side in teal refers to the prices of all hours in which the unintended storage cycling occurs. The bubbles represent underlying discrete prices to visualize accumulations and outliers. The numbers at the top indicate average prices, while the numbers at the bottom refer to the number of hourly prices for each of the violin's sides.}
\label{fig:violine}
\end{figure}

Figure~\ref{fig:violine} illustrates the estimated distribution (kernel density estimation) of prices,\footnote{We interpret the dual variables of the energy balance as hourly wholesale prices, compare \cite{brown_decreasing_2021} for model specifications based on constraint family (1).} which vary across \ac{SLCR} levels for two reasons. First, as the capacity portfolio changes across \ac{SLCR} levels, the shape of the merit order and the intersection of the supply curve with demand may be affected. Second, varying optimality conditions for storage charging and discharging drive the differences in wholesale prices between different \ac{SLCR} levels. 

Figure ~\ref{fig:violine} also shows that unintended storage cycling occurs only in low-price periods, i.e.,~in hours with renewable curtailment. Note that prices may become negative in models using constraint family (1), which can be interpreted as the consequence of an energy-based renewable support mechanism \cite{brown_decreasing_2021,prol_economics_2020}.

Price formations of models based on constraint families (2) to (4) are characterized by the same variation in the distribution across \ac{SLCR} levels, but a constant level shift may apply for each constraint family. This effect is due to a variation of optimality conditions of generators across constraint families. It may be explored in more detail in the future, but is beyond the scope of this paper.

\subsection{Drivers of unintended storage cycling}\label{ssec:drivers_storage_cycling}

To disentangle the drivers of unintended storage cycling, we conduct sensitivity analyses with respect to key input parameters: the renewable energy target, storage efficiency, as well as the variable costs of storage use, renewable generation, and renewable curtailment.

\begin{figure}[htbp]
\centering
\subfloat[Different renewable penetration levels (no occurrence below 40\%).\label{fig:storage_cycling_phi_e}]
    {\includegraphics[width=0.47\textwidth]{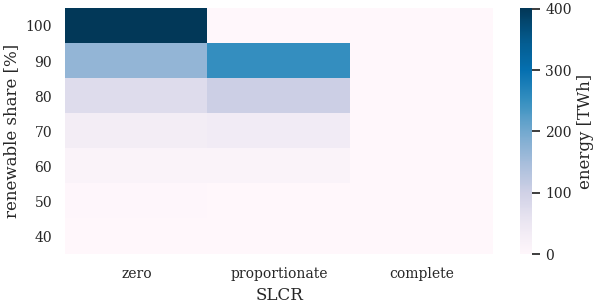}}
\qquad
\subfloat[Different levels of round-trip storage efficiency.\label{fig:storage_cycling_eta_e}]
    {\includegraphics[width=0.47\textwidth]{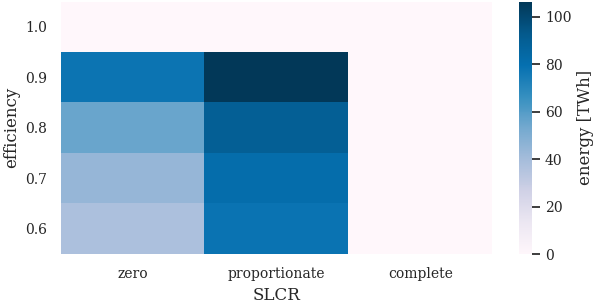}}\\
\subfloat[Different levels of variable costs of storage use.\label{fig:storage_cycling_cvar_sto_e}]
    {\includegraphics[width=0.47\textwidth]{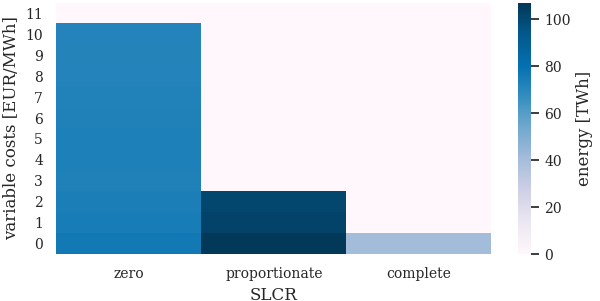}}
\qquad
\subfloat[Different levels of variable costs of renewable generation (no occurrence above 9000~EUR/MWh for zero \ac{SLCR}).\label{fig:storage_cycling_cvar_res_e}]
    {\includegraphics[width=0.47\textwidth]{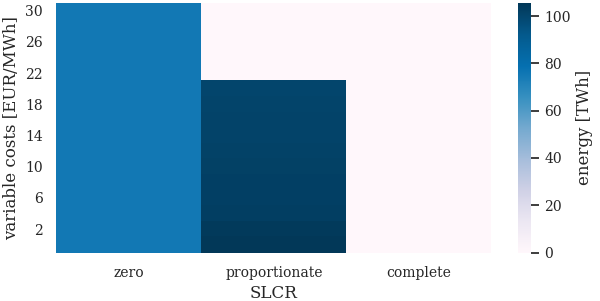}}
\qquad
\subfloat[Different levels of variable costs of renewable curtailment.\label{fig:storage_cycling_cvar_cu_e}]
    {\includegraphics[width=0.47\textwidth]{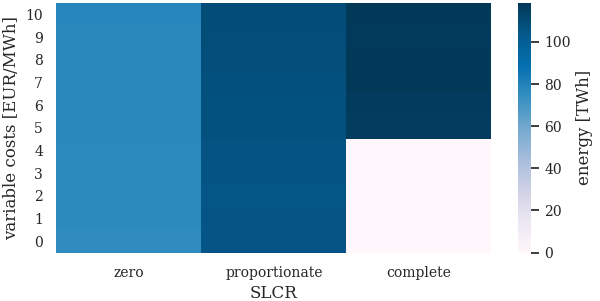}}\\
\caption{Sensitivity analysis of the annual energy of unintended storage cycling for different \ac{SLCR} levels. A renewable share of 80\% applies to all panels except panel \ref{fig:storage_cycling_phi_e}. Despite marginal discrepancies, results coincide across all constraint families.}
\label{fig:sto_cyc_sensitivity}
\end{figure}

\begin{figure}[htbp]%
\centering
\subfloat[Energy flows at a renewable share of 80\%.\label{fig:sankey_80_1a}]
    {{\includegraphics[width=\textwidth]{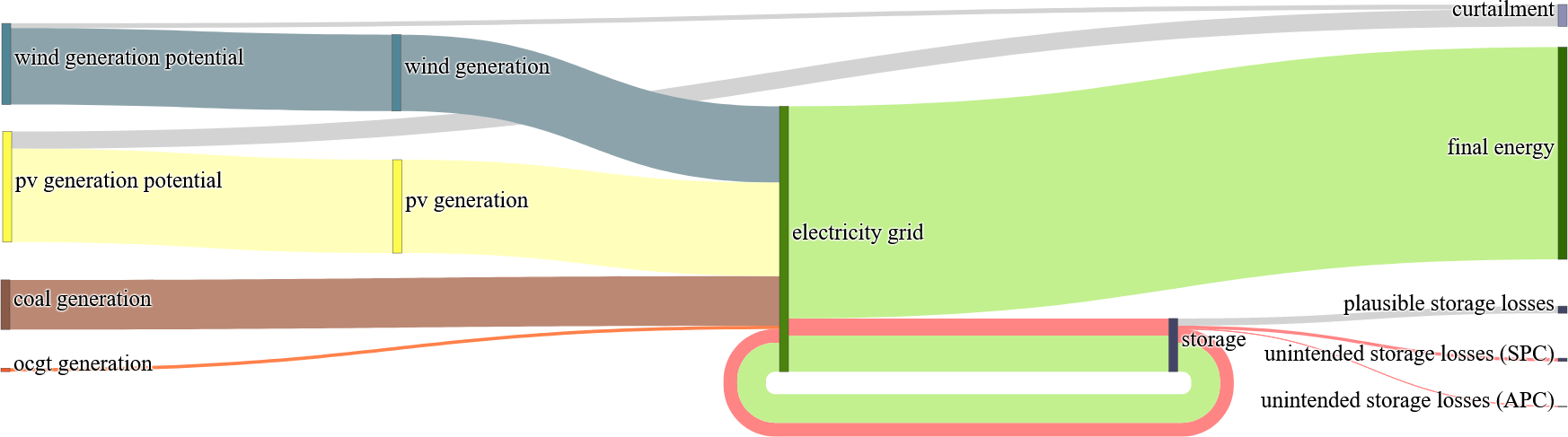}}}\\
\subfloat[Energy flows at a renewable share of 100\%. \label{fig:sankey_100_1a}]
    {{\includegraphics[width=\textwidth]{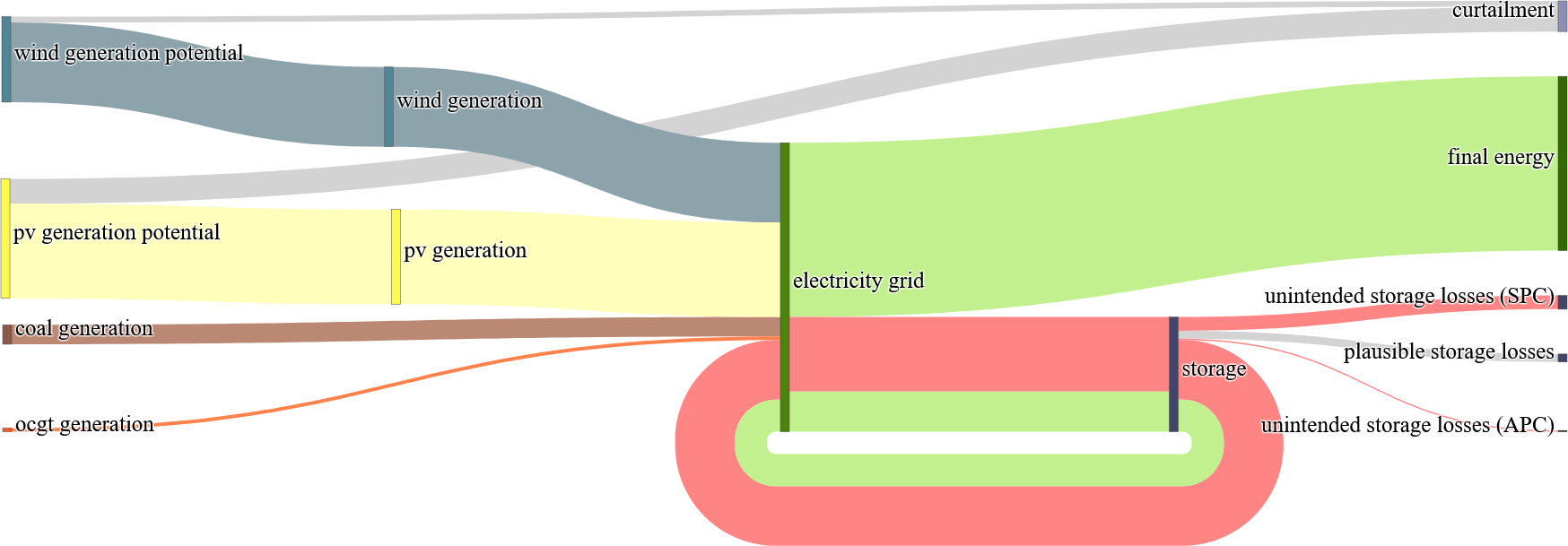}}}\\
\caption{Energy flows for a power sector based on the model (1a) using a renewable share constraint with zero \ac{SLCR}. The height of each trace represents the energy amount. Light green traces represent intended power flows withdrawn from or fed back into the grid. Red traces indicate unintended power flows due to unintended storage cycling. Unintended storage cycling and unintended storage losses grow with the renewable share target.}%
\label{fig:sankey}%
\end{figure}

With increasing renewable penetration, the number of hours with unintended storage cycling and the energy related to the artifact grows disproportionately (Figure~\ref{fig:storage_cycling_phi_e} and \ref{fig:sankey}). Below a renewable share of 40\%, no storage is deployed due to very limited \ac{VRE} surplus generation. Accordingly, unintended storage cycling does not occur. As the required \ac{VRE} penetration increases, \ac{VRE} surplus energy and optimal electricity storage capacities also grow. Unintended storage cycling becomes available and is increasingly used. In a fully renewable power sector, model specifications with proportionate and complete \ac{SLCR} converge, since the renewable share constraints coincide (see Table \ref{tab:res_shares} for $\phi = 1$). Both of these \ac{SLCR} levels then prevent unintended storage cycling altogether (see top row in Figure~\ref{fig:storage_cycling_phi_e}), while model specifications with zero \ac{SLCR} still suffer from the artifact (see also Figure \ref{fig:sankey_100_1a}). By construction, the latter do not comply with the notion of a fully renewable power sector even at a renewable share of 100\%, as they do not explicitly require complete storage losses to be covered by renewables. Instead, some level of conventional generation always remains in the system under zero \ac{SLCR} constraint formulations.

The round-trip efficiency of the electricity storage technology determines storage losses. 
Yet, unintended storage cycling does not increase with lower round-trip efficiency (below a value of 1.0). Instead, \ac{LCOS} decline as efficiency increases. The system's benefit of storage grows, and yields higher optimal storage capacity. This enables higher levels of unintended storage cycling (Figure~\ref{fig:storage_cycling_eta_e}). A round-trip efficiency of~1.0, a strong (and unrealistic) assumption for electricity storage, averts storage losses and unintended storage cycling altogether.

Increasing variable costs of storage use, attributed to each unit of charging and discharging, are another driver of unintended storage cycling (Figure~\ref{fig:storage_cycling_cvar_sto_e}). If these costs are zero, the economic trade-off between curtailment of \ac{VRE} surplus and unintended storage losses cannot be quantified. The optimization becomes indifferent between the two options, and unintended storage cycling may occur at undetermined levels. As variable cost for storage use rise, the cost advantage of removing \ac{VRE} surplus via unintended storage cycling declines. Above a certain threshold, which is substantially higher than the variable costs of current storage technologies \cite{lacal_arantegui_etri_2014}, unintended storage cycling becomes an unfavorable option in the cost minimization. In our stylized setting, this threshold is much lower for models with proportionate \ac{SLCR} than for models with zero \ac{SLCR}.

The same holds true for increasing variable costs of \ac{VRE} generation, which incentivize curtailment of \ac{VRE} surplus rather than feeding it into the grid. This makes unintended storage cycling less favorable, as converting renewable surplus into additional storage losses increases the variable cost of renewable generation. Above a certain threshold, the costs of unintended storage cycling thus surpass its gains, and the artifact disappears (Figure~\ref{fig:storage_cycling_cvar_res_e}). Yet, this threshold is beyond of what appears plausible for real-world \ac{VRE} technologies \cite{lacal_arantegui_etri_2014}.

All results presented so far were generated under the assumption that \ac{VRE} curtailment comes at no cost. Rising variable costs of curtailment
leads to decreasing curtailment levels. Instead, \ac{VRE} generation at times of high \ac{VRE} availability increases, which is then removed from the system via additional unintended storage cycling (Figure~\ref{fig:storage_cycling_cvar_cu_e}) in settings with incomplete \ac{SLCR}. At very high curtailment costs this becomes effective even for model specifications with complete \ac{SLCR}. 

\section{Discussion}\label{sec:discussion}

\subsection{A remedy for unintended storage cycling using renewable energy constraints}\label{ssec:solutions_res1}

We have shown that unintended storage cycling can cause significant disruptions of optimal dispatch and investment patterns in models that include a constraint on annual renewable energy generation. To remedy this artifact, the substitution of \ac{VRE} curtailment with additional storage losses can be prohibited by fully including storage losses to the renewable energy constraint. This is done in all model specifications with complete \ac{SLCR}, regardless of the constraint family.

Real-world renewable energy targets are often defined as the ratio of annual generation from renewable energy sources over total electricity generation (or total demand) in the system. This implies that storage losses should also be distributed among renewable and conventional generators according to the targeted renewable share. Model specifications with proportionate \ac{SLCR} conform with this notion, thus appearing as most policy-relevant. In contrast, models with complete \ac{SLCR} are overly restrictive in the sense that they require additional renewable generation to completely cover storage losses.

Notwithstanding, we recommend applying a renewable energy constraint with complete \ac{SLCR}, as this remedies unintended storage cycling. In post-optimization, the achieved renewable share could still be reported according to both complete and proportionate \ac{SLCR}. The latter can then be interpreted and communicated as the actual renewable share of the system, which is slightly higher than the former. For example, in our model setting with complete \ac{SLCR} and a renewable target of 80\%, the actual renewable share reported according to the formulation with proportionate \ac{SLCR} is 81.16\%. We consider such a slightly increased renewable share to be a reasonable price for energy modelers to pay to avoid unintended storage cycling, especially as renewable shares are set to increase anyway in light of ongoing ambitions to achieve carbon-neutrality. Moreover, models with proportionate and complete \ac{SLCR} converge in a fully renewable setting, which, most likely, is required for carbon-neutrality. 

Alternatively, modelers may attempt to slightly decrease the desired renewable target in models with complete \ac{SLCR} to achieve a system configuration that meets the desired target reported as per the proportionate \ac{SLCR} model. For example, we need to specify a renewable share of 79.3\% in a model with complete \ac{SLCR} in order to achieve an actual renewable energy share of 80\% reported according to a formulation with proportionate \ac{SLCR}. However, the identification of the required share as per complete \ac{SLCR} would require trial-and-error model runs that need to be done for each model parameterization individually. Depending on the scope of the model and its computational burden, this may be infeasible because of significant modeling time and effort, as well as a high number of trails.

Based on the results shown in Section~\ref{ssec:drivers_storage_cycling}, one option to avoid over-restricting energy models with complete \ac{SLCR} would be to increase the variable costs of storage use and renewable generation. This makes unintended storage cycling less favorable in the sense of the objective function and may allow using a more intuitive constraint formulation with proportionate \ac{SLCR}. However, this strategy has to be taken with caution. In the real world, variable storage costs are generally low, irrespective of the technology. A manipulation of these costs distorts the relative costs of the available technology portfolio, and thus also optimal dispatch and investment decisions. The deployment of storage may decline, potentially causing far-reaching system-wide effects, such as a hindered integration of \ac{VRE}. Moreover, identifying an appropriate increase in variable storage costs that does not substantially affect optimal storage deployment while preventing unintended storage cycling is not straightforward. Instead, it requires extensive testing, and may even be infeasible, depending on the model parameterization. We thus advise against the deliberate manipulation of storage cost assumptions to remedy unintended storage cycling when using renewable energy targets.

\subsection{Potential remedies for unintended storage cycling using alternative renewable targets}\label{ssec:solutions_res2}

There are two alternative options for avoiding unintended storage cycling while using renewable targets with a basis other than annual energy. First, instead of using the actual renewable energy generation (after curtailment), the theoretical renewable generation potential (before curtailment) could serve as the base element in the renewable energy constraint (see Equation~(\ref{eq:general_RES_con})) \cite{gils_integrated_2017,scholz_application_2017,child_2019,bogdanov_2019}. Relaxing the renewable energy constraint via unintended storage cycling is then impossible, since the substitution of curtailment with additional storage losses cannot contribute useful energy to the renewable target. However, this approach changes the nature of the renewable energy target. Models constrained this way optimize the system such that there is sufficient \ac{VRE} capacity to achieve some renewable energy generation \textit{potential} relative to total demand or total generation. Yet, actual renewable generation is not constrained and may be much lower. Modeling scenarios that adhere to binding renewable energy policy targets in terms of actual renewable generation is thus impossible. Furthermore, the composition of the optimal technology portfolio may significantly deviate from models with complete \ac{SLCR}, as the role of conventional generators could be overestimated, while the need for renewable generation capacity and appropriate balancing options may be underestimated.

Second, an exogenous expansion target for renewable capacity could be used \cite{pfenninger_renewables_2015,hirth_market_2013,anke_expansion_2021}. This approach foresees a capacity mix with an exogenous renewable part, while the remaining part is determined endogenously. Since the actual generation of each technology is not optimized to realize a certain share of renewable energy, the latter may be much lower than desired. Due to this uncertainty of actual renewable generation, binding renewable energy targets cannot be investigated. This could be resolved by a manual determination of exogenous renewable capacity targets to achieve a desired renewable energy target. However, identifying concrete renewable capacities \textit{a priori} is not straightforward and may require a large number of trials. This becomes particularly relevant for comprehensive model settings that include multiple sector coupling technologies, which puts the feasibility of this approach into question.

\subsection{Potential remedies for unintended storage cycling using emission constraints}\label{ssec:solutions_emmissions}

Instead of including renewable targets, modelers may constrain or penalize carbon emissions to achieve high renewable shares and, at the same time, avoid unintended storage cycling. Based on our literature review, we discuss two options: a limitation of the allowable emission budget (carbon cap), and an explicit carbon price. 

Given $e_s$ being the specific emissions per generator $s$ and $CB$ the available carbon budget, a constraint implementing a carbon cap could be formulated as:

\begin{equation}\label{eq:carbon_cap}
    \sum\limits_{s,t} e_s G_{s,t} \leq CB
\end{equation}

The dual variable of this constraint indicates how much total system costs change if the carbon cap tightens by one marginal unit. We interpret this as the system costs for one unit of emission reduction. In a long-term equilibrium, this implicit carbon price can be interpreted as a carbon price paid by emitting generators \cite{brown_decreasing_2021}. Note that setting policy-relevant carbon budgets in electricity models may be challenging, as political targets for the power sector often do not exist.

In contrast, setting an explicit carbon price on each unit of emissions is more easy to implement, and it becomes directly effective as an additional variable cost component in the objective function.

In both settings, unintended storage cycling cannot occur since renewable energy generation is not constrained. The conversion of \ac{VRE} curtailment into additional storage losses thus never leads to lower-cost solutions. Importantly, this only holds true if variable costs for storage use are greater zero. Otherwise, the model may be indifferent between \ac{VRE} curtailment and storage losses.

In general, outcomes of models with similar renewable shares may substantially differ, depending on what drives this share, a carbon constraint, or a renewable constraint. The latter promotes additional renewable generation or capacity beyond the unconstrained optimum. This is done by introducing an energy- or capacity-oriented subsidy for renewable generators, which increases their revenues. In contrast, carbon policy instruments impact the use of conventional generators by increasing their costs. This forces a decline especially of those generators with the highest carbon emissions, which may be replaced by lower-emission technologies. The resulting capacity and energy mix may significantly differ from the solution based on a binding renewable target. A binding carbon cap favors lower-emission technologies (e.g.,~gas-fired plants) over higher-emission technologies (e.g.,~lignite-fired plants). As the former tend to have higher variable and lower capital costs, this shift may decrease the optimal storage deployment and affect the overall capacity mix.

Explicit carbon pricing affects the merit order of available generation technologies, increasing the variable cost of emitting technologies. Neither renewable energy nor carbon emissions are constrained. The resulting optimal capacity and energy mix resembles the unconstrained optimum that reflects the economics of the available technology portfolio. Depending on the costs and availability of flexibility options in the model, substantial carbon prices may be required to achieve very high shares of renewables. Moreover, the determination of appropriate price levels \textit{a priori} may be difficult since not only economic but also political, environmental, and social aspects may impact price formations on the electricity wholesale market.

While model formulations that constrain or penalize carbon emissions can effectively avoid unintended storage cycling, they generally lead to different outcomes than models with explicit renewable energy targets. Hence, they are not an adequate substitute for cases where the effects of explicit renewable targets are of interest, e.g., in many countries that have set such renewable targets.

\subsection{Unintended energy losses in settings with sector coupling}

The use of renewable electricity for mobility, heating, and industrial processes is considered a main strategy for decarbonizing the energy system \cite{de_coninck_global_2018}. This strategy is often referred to as sector coupling. It also includes indirect electrification via green hydrogen, which may be used where direct electrification is not viable \cite{jacobson_100_2017,hanley_2018}. 

The mechanism of unintended storage cycling described here for electricity storage may also apply to various sector coupling technologies, as these also come with inherent energy losses. Here, the model artifact may materialize in the form of \ac{VRE} surplus being converted into unintended energy losses.

There are three relevant types of such losses: First, losses inherent to sector coupling options that allow for re-conversion to electricity, i.e.,~with an operation principle similar to electricity storage. These take up energy from the power sector and convert it to some intermediate form. The energy is then either consumed for end energy use in some other sector, or, if beneficial from the system perspective, partly fed back into the grid at a later point in time. Examples are battery-electric vehicles with a vehicle-to-grid discharging option, or power plants fueled with green hydrogen.\footnote{The same applies to load-shifting activities that incur energy losses.} To remedy unintended energy losses, we recommend to adjust the loss term of the renewable energy constraint such that all relevant losses are covered by renewable generators. Only losses related to processes enabling unintended energy cycling are relevant in this context (e.g.,~losses related to discharging electricity back from an electric vehicle battery to the grid). Losses from energy consumption without re-conversion option (e.g.,~losses related to actually driving an electric vehicle) are not relevant here, if the respective demand of a sector coupling option is exogenous to the model (see Section \ref{ssec:usc_remedy_sector_soupling}).

Second, there are standing losses through self-discharge. These may occur in various types of energy storage, e.g., in different types of low- and high-temperature heat storage, or in liquid hydrogen storage.\footnote{Standing losses may also be relevant for some long-duration electricity storage technologies.} The model artifact may reoccur here in the form of distorted storage use patterns that aim at increasing standing losses by stretching the storage period. Again, properly accounting for standing losses in the loss term of the renewable energy constraint would remedy the issue.

Third, there may be unintended losses arising from sub-optimal technology choices. Suppose there are two different sector coupling options available, one of them more energy-efficient than the other one. Optimal use and investment decisions may then skew away from the alternative with lower losses towards the one with higher losses. This would allow more \ac{VRE} surplus from the power sector to be converted into energy losses that relax the renewable constraint. An example for such alternative technologies are different supply chains of green hydrogen, such as liquid or gaseous hydrogen \citep{stockl_optimal_2021}. We leave it to future research to find a remedy for distortions related to this type of unintended energy losses.

\subsection{Unintended energy losses in multi-regional settings with transmission losses}

Electricity trade facilitates spatial smoothing of \ac{VRE} generation and electric load in multi-regional settings, especially in the absence of other flexibility options \cite{brown_synergies_2018}. This causes losses in the transmission grid due to the resistance of power lines. Unintended and excessive shifts of energy across and between nodes or regions may arise to generate additional grid losses that remove \ac{VRE} surplus from the system instead of curtailing it at its place of origin. The additional \ac{VRE} generation would then help to relax the renewable energy constraint, which lowers the need for \ac{VRE} capacity and renders a lower-cost solution possible. However, the exported energy does not or only partly serve demand in the importing region, and generation from other generators needs to serve demand. Unintended transmission losses may thus affect optimal use and investment patterns of generation and transmission infrastructure as well as their spatial allocation across regions. 

A model specification with a system-wide renewable share referenced to either total demand or generation accumulated across all regions is particularly prone to unintended transmission losses. This is because the spatial allocation of generation and transmission capacity is not constrained, but freely relocatable across all regions. Optimal \ac{VRE} generation capacity may increase in regions that have low \ac{VRE} \ac{LCOE} due to superior weather conditions. At least some of the additional renewable surplus could be converted into additional renewable generation here, and, subsequently, removed from the system in form of unintended transmission losses. This decreases the need for \ac{VRE} capacity installations in regions with high \ac{LCOE} related to inferior weather conditions. Instead, dispatchable technologies with lower \ac{LCOE} can serve demand here. This capacity shift across regions may lead to lower-cost solution than settings without unintended transmission losses. To prevent the artifact from occurring, the renewable energy constraint would have to require renewable generators to completely cover all transmission losses arising across all regions.

Unintended transmission losses may also arise in settings with disaggregated renewable targets specific to each region. A capacity and portfolio effect similar to the one caused by unintended storage cycling may arise here, only that \ac{VRE} surplus is exported and partly converted into unintended transmission losses instead of unintended storage losses. Again, we recommend to add transmission losses to the loss term of the renewable energy constraint. Note that total domestic demand does not include the energy that is lost in the transmission grid. Referenced to domestic demand, the loss term of the renewable energy constraint should thus include total losses associated with imports required to supply domestic demand. In contrast, total domestic generation includes the energy lost in the transmission grid when exporting to other countries. When referencing the domestic renewable energy constraint to some fraction of domestic generation, only a fraction of the transmission losses from exports to other regions are considered. The remainder should be added to the loss term of the renewable energy constraint (cp. the treatment of unintended storage losses in the loss term of constraint (2c) in Table \ref{tab:res_shares}). Again, this merits further investigation in future research.

\section{Summary and conclusion}\label{sec:summary_conclusion}

In this paper, we describe and investigate the modeling artifact of unintended storage cycling, which may arise in a wide range of cost-minimizing energy models that make use of binding renewable energy constraints. In such models, it can be beneficial not to curtail renewable surplus energy, but to convert excess electricity into additional storage losses. This artifact can be detected when there are periods with simultaneous storage charging and discharging. The respective increase in renewable generation can be realized without additional renewable capacity installations, thus helping to achieve the renewable energy target at lower costs. This may distort optimal dispatch and investment decisions of all technologies. 

We specifically explore the impact of different constraint formulations for renewable energy targets on model outcomes and unintended storage cycling. Using a parsimonious power sector model, we analytically derive intuition for the cause of unintended storage cycling from an energy modeler's perspective. We further investigate how including storage losses in the renewable energy constraint affects optimality conditions for storage in the long-term equilibrium. Only if storage losses have to be completely covered by renewable generation, unintended storage cycling can be prevented. Loosely parameterizing the model to the German power sector, we show that the artifact can substantially distort optimal dispatch and investment decisions as well as modeled market prices.

To prevent unintended storage cycling, we recommend to completely cover storage losses in the renewable energy constraint, if such a constraint is used in energy models. This comes at the expense of a slight over-restriction of the model concerning the renewable target, but we argue that this is a reasonable price to pay to avoid the artifact. Furthermore, this can be addressed by adjusted reporting and target setting. Alternatively, other types of renewable constraints not based on energy could be used to avoid the problem, but these come with other drawbacks. Likewise, the artifact can be resolved by implementing carbon emission constraints or penalties. Yet, as these change overall power sector outcomes, they may not be suitable for some types of policy-relevant analyses where binding renewable energy targets play a role.

Unintended storage cycling may be considered a special case of a more general phenomenon of unintended energy losses. It may also occur in models with sector coupling technologies and transmission. Here, various types of energy losses could be traded-off against renewable curtailment to relax the renewable energy constraint in least-cost model solutions. In general, including such losses in the renewable constraint may help in many cases. We leave it to future research to explore the feasibility and the limits of this approach for complex models with many sector coupling technologies and transmission.

\section*{Acknowledgments}
We thank Mario Kendziorski and Alexander Zerrahn for valuable inputs and discussions, as well as Alexander Roth and Mariza Montes de Oca León for feedback on earlier drafts. We also thank the participants of various conferences and seminars for helpful comments, including the DIW Berlin online workshop ``The role of energy storage in power sectors with fossil fuel phase out'' on 9 October 2020, the PhD seminar ``Economics of Fossil Fuel Phase-out'' on 20 November 2020, the International Renewable Energy Storage (IRES) Conference 2021, the International Conference on Energy Economics and Technology (Enerday) 2021, the Young Energy Economists and Engineers Seminar (YEEES) 2021, the International Association for Energy Economics (IAEE) Online Conference 2021, and the Annual Conference of the European Association of Environmental and Resource Economists (EAERE) 2021. We acknowledge funding by the German Federal Ministry of Education and Research via the project ``Future of Fossil Fuels in the wake of greenhouse gas neutrality'' (grant number FKZ 01LA1810B), and by the German Federal Ministry for Economic Affairs and Energy via the project ``MODEZEEN'' (grant number FKZ 03EI1019D).

\section*{Author Contributions} \label{sec:author contributions}


\textbf{Martin Kittel}: Conceptualization, methodology, software, formal analysis, investigation, data curation, visualization, writing - original draft, writing - review and editing, supervision, project administration. \textbf{Wolf-Peter Schill}: Conceptualization, investigation, data curation, writing - review and editing, project administration.

\section*{}

\bibliography{bib_no_sync}

\newpage

\renewcommand{\thesection}{SI}
\global\long\def\thefigure{SI.\arabic{figure}}
\setcounter{figure}{0}

\global\long\def\thetable{SI.\arabic{table}}
\setcounter{table}{0}

\global\long\def\theequation{SI.\arabic{equation}}
\setcounter{equation}{0}

\section{Supplementary information}\label{sec:sup_inf}

\subsection{Constraint formulations}\label{sec:constraints_table}

\begin{table}[htbp]
\centering
\ra{1.1}
\caption{Minimum renewable and maximum conventional shares considered in the optimization problem discussed in Section \ref{ssec:model_definition}.}
\label{tab:res_shares_opt}
\vspace{.1cm}
\small
\begin{tabular}{lclr}\toprule
No. & \ac{SLCR} & share & constraint \\
\midrule
1a & zero & $\Omega = \phi \sum\limits_t d_t$ & (\ref{eq:theta}) \\
1b & proportionate & $\Omega = \phi \Big( \sum\limits_t d_t +\sum\limits_{r,t} (G^{in}_{r,t} - G^{out}_{r,t}) \Big)$ & (\ref{eq:theta}) \\
1c & complete & $\Omega = \phi \sum\limits_t d_t + \sum\limits_{r,t} (G^{in}_{r,t} - G^{out}_{r,t})$ & (\ref{eq:theta}) \\
\midrule
2a & zero & $\Omega = \phi \Big( \sum\limits_{s,t} G_{s,t} - \sum\limits_{r,t} (G^{in}_{r,t} - G^{out}_{r,t}) \Big)$ & (\ref{eq:theta}) \\
2b & proportionate & $\Omega = \phi \sum\limits_{s,t} G_{s,t}$ & (\ref{eq:theta}) \\
2c & complete & $\Omega = \phi \sum\limits_{s,t} G_{s,t} + (1-\phi) \sum\limits_{r,t} (G^{in}_{r,t} - G^{out}_{r,t})$ & (\ref{eq:theta})\\
\midrule
3a & zero & $\Theta = (1 - \phi) \sum\limits_t d_t + \sum\limits_{r,t} (G^{in}_{r,t} - G^{out}_{r,t})$ & (\ref{eq:omega}) \\
3b & proportionate & $\Theta = (1 - \phi) \Big( \sum\limits_t d_t + \sum\limits_{r,t} (G^{in}_{r,t} - G^{out}_{r,t}) \Big)$ & (\ref{eq:omega}) \\
3c & complete & $\Theta = (1 - \phi) \sum\limits_t d_t$ & (\ref{eq:omega}) \\
\midrule
4a & zero & $\Theta = (1 - \phi) \sum\limits_{s,t} G_{s,t} + \phi \sum\limits_{r,t} (G^{in}_{r,t} - G^{out}_{r,t})$ & (\ref{eq:omega}) \\
4b & proportionate & $\Theta = (1 - \phi) \sum\limits_{s,t} G_{s,t}$ & (\ref{eq:omega}) \\
4c & complete & $\Theta = (1 - \phi) \Big( \sum\limits_{s,t} G_{s,t} - \sum\limits_{r,t} (G^{in}_{r,t} - G^{out}_{r,t}) \Big)$  & (\ref{eq:omega}) \\
\bottomrule
\end{tabular}
\end{table}

\subsection{Cost assumptions}\label{sec:cost_table}

\begin{table}[h!]
\centering
\ra{1.1}
\caption{Overview of assumptions on investment, annual fixed and variable costs.}
\label{tab:cost_assumptions}
\vspace{.1cm}
\small
\begin{tabular}{lccr}\toprule
 & overnight investment costs & annual fixed costs & variable costs \\
technology & [EUR/kW(h)] & [EUR/kW] & [EUR/MWh] \\
\midrule
coal & 1300 & 25 & 21.55\\
ocgt & 400 & 1.5 & 76.34 \\
pv & 390 & 10.6 & - \\
wind & 1000 & 20 & - \\
storage charging & 1.1 & - & 0.5 \\
storage discharging & 1.1 & - & 0.5 \\
storage energy & 80 & - & - \\
\bottomrule
\end{tabular}
\end{table}

\subsection{Mathematical background - Lagrangian function}\label{sec:lagrange}

Suppose that $f$, $g_l$, and $h_k$ are affine, continuously differentiable functions. Our constrained optimization problem minimizes the objective function $f$ over decision variables $x_i$, subject to equality constraints $g_l$ and inequality constraints $h_k$:

\begin{mini}|l|
    {x_i}{f(x_i)}{}{}
    \label{eq:generalLP}
	\addConstraint{g_l(x_i)}{=0 \perp \lambda_l}{\quad \forall l}
	\addConstraint{h_k(x_i)}{\geq 0 \perp \mu_k}{\quad \forall k}
\end{mini}
\vspace{.1cm}

For optimality of a minimization problem with the inequality sign convention as in Equation (\ref{eq:generalLP}), we require parallel objective and constraint gradients at the tangential point \cite[Ch. 18.5]{simon_mathematics_1994}:

\begin{equation}
    \nabla f(x_i) = \sum_l \lambda_l \nabla_x g_l(x_i) + \sum_k \mu_k \nabla_x h_k(x_i)
\end{equation}
\vspace{.1cm}

Based on the optimality condition and the problem set-up in Equation (\ref{eq:generalLP}), we establish the Lagrangian function with non-negative multipliers of inequality constraints \cite[Ch. 18.5]{simon_mathematics_1994}:

\begin{equation}
    \mathcal{L}(x_i, \lambda_l, \mu_k) = f(x_i) - \sum_l \lambda_l g_l(x_i) - \sum_k \mu_k h_k(x_i)
\end{equation}
\vspace{.1cm}

With this formulation, $\mu_k$ indicates the change in the objective if $h_k(x_i)$ is relaxed. For instance, suppose $h_k(x_i)$ includes a constant $c$, such that $h_k(x_i) \leq c$. If $c$ marginally increased, that is $h_k(x_i) \leq c + \epsilon$, $\mu_k$ represents the corresponding rise in $f(x_i)$. In the optimum, which is characterized by the vector $(x_i^*, \lambda_l^*, \mu_k^*)$, the following four \ac{KKT} conditions are satisfied. First, stationarity, i.e.,~there is no feasible direction to improve the objective:

\begin{equation}
    \nabla_x \mathcal{L}(x_i^*, \lambda_l^*, \mu_k^*) = \nabla_x f(x_i^*) - \sum_l \lambda_l^* \nabla_x g_l(x_i^*) - \sum_k \mu_k^* \nabla_x h_k(x_i^*) = 0
\end{equation}

\noindent Second, primal feasibility ensures feasibility of the (in-)equality constraints:

\begin{subequations}
    \begin{align}
        g_l(x_i^*) &= 0\\
        h_k(x_i^*) &\leq 0
    \end{align}
\end{subequations}

Third, while the multipliers (also called dual variables) of the equality constraints are free, dual feasibility establishes non-negative multipliers for inequality constraints:

\begin{equation}
    \mu_k^* \geq 0
\end{equation}

Finally, complementary slackness conditions require a positive multiplier $\mu_k^* > 0$ if an inequality constraint is binding, i.e.,~$h_k(x_i^*) = 0$, or $\mu_k^* = 0$ if the constraint is not binding, i.e.,~$h_k(x_i^*) < 0$:

\begin{equation}
    \mu_k^* h_k(x_i^*) = 0
\end{equation}

\subsection{Derivation of optimality conditions in an unconstrained optimum}\label{ssec:derivations_unconstrained_opt}

In the unconstrained optimum of the model introduced in Section \ref{ssec:model_definition}, renewable energy constraints (\ref{eq:omega}) and (\ref{eq:theta}) do not apply. In this case, \ac{KKT} stationarity yields the following first-order conditions for storage:

\begin{align}
        \frac{\partial \mathcal{L}}{\partial C_r^\circ} &= 0 \Rightarrow i_r^\circ - \sum_t \bar{\mu}_{s,t}^\circ = 0 \label{eq:kkt_C_r_unconstr}\\
        \frac{\partial \mathcal{L}}{\partial G^{out}_{r,t}} &= 0 \Rightarrow o^{out}_r - \lambda_t - \ubar{\mu}^{out}_{r,t} + \bar{\mu}^{out}_{r,t} - (\eta^{out}_r)^{-1} \lambda^l_{r,t} = 0 \label{eq:kkt_G_r_out_unconstr} \\
        \frac{\partial \mathcal{L}}{\partial G^{in}_{r,t}} &= 0 \Rightarrow o^{in}_r + \lambda_t - \ubar{\mu}^{in}_{r,t} + \bar{\mu}^{in}_{r,t} + \eta^{in}_r \lambda^l_{r,t} = 0 \label{eq:kkt_G_r_in_unconstr}\\
        \frac{\partial \mathcal{L}}{\partial G^{l}_{r,t}} &= 0 \Rightarrow - \ubar{\mu}^{l}_{r,t} + \bar{\mu}^{l}_{r,t} - \lambda^l_{r,t} + \lambda^l_{r,t+1} = 0 \label{eq:kkt_G_r_l_unconstr}
\end{align}

\noindent \ac{KKT} complementary slackness requires the following for the storage inequality constraints:

\begin{align}
    \ubar{\mu}^\circ_{r,t} G^\circ_{r,t} &= 0 \label{eq:slackness_r_lower_bound_unconstr}\\
    \ubar{\mu}^\circ_{r,t} (C_r^\circ - G^\circ_{r,t}) &= 0 \label{eq:slackness_r_upper_bound_unconstr}
\end{align}

\noindent We establish the zero-profit rule for storage of the unconstrained optimum as follows:

\begin{equation}\label{eq:derivation_zpr_r_unconstr}
\begin{split}
    \sum_\circ i_r^\circ C_r^\circ &+ \sum_{t,*} o_r^* G_{r,t}^* + \sum_t \lambda_t G_{r,t}^{in}  \\
    & = \sum_{t,\circ} \bar{\mu}^\circ_{r,t} C^\circ_r + \sum_t G_{r,t}^{out} \Big(\lambda_t + \ubar{\mu}^{out}_{r,t} - \bar{\mu}^{out}_{r,t} + \lambda^l_{r,t} (\eta^{out}_r)^{-1} \Big) \\
    & \quad + \sum_t G_{r,t}^{in} \Big(- \lambda_t + \ubar{\mu}^{in}_{r,t} - \bar{\mu}^{in}_{r,t} - \lambda^l_{r,t} \eta^{in}_r \Big) + \sum_t \lambda_t G_{r,t}^{in} \\
    & = \sum_{t,\circ} \bar{\mu}^\circ_{r,t} C^\circ_r + \sum_t G_{r,t}^{out} \Big(\lambda_t + \ubar{\mu}^{out}_{r,t} - \bar{\mu}^{out}_{r,t} + \lambda^l_{r,t} (\eta^{out}_r)^{-1} \Big) \\
    & \quad + \sum_t G_{r,t}^{in} \Big(- \lambda_t + \lambda_t + \ubar{\mu}^{in}_{r,t} - \bar{\mu}^{in}_{r,t} - \lambda^l_{r,t} \eta^{in}_r \Big) \\
    & = \sum_t \Big( \ubar{\mu}^{out}_{rt} G^{out}_{r,t} + \ubar{\mu}^{in}_{rt} G^{in}_{r,t} + \bar{\mu}^{out}_{rt} (C^{out}_r - G^{out}_{r,t}) + \bar{\mu}^{in}_{rt} (C^{in}_r - G^{in}_{r,t}) \\
    & \quad + G^{out}_{r,t} \big(\lambda_t + \lambda^l_{r,t} (\eta^{out}_r)^{-1} \big) + G^{in}_{r,t} \big(- \lambda^l_{r,t} \eta^{in}_r \big) + G^l_{r,t} \big( \ubar{\mu}^l_{r,t} + \lambda^l_{r,t} - \lambda^l_{r,t+1} \big) \Big) \\
    &= \sum_t \Big( \lambda_t G_{r,t}^{out} + \ubar{\mu}^l_{r,t} G^l_{r,t} + \lambda^l_{r,t} \big(G^l_{r,t} - \eta_r^l G^l_{r,t-1} - \eta^{in}_r G^{in}_{r,t} + (\eta^{out}_r)^{-1} G^{out}_{r,t} \big) \Big)\\
    &= \sum_t \lambda_t G_{r,t}^{out}
\end{split}
\end{equation}
\vspace{0.1cm}

The first step employs stationarity conditions (\ref{eq:kkt_C_r_unconstr}), (\ref{eq:kkt_G_r_out_unconstr}) and (\ref{eq:kkt_G_r_in_unconstr}). The second step rearranges terms, uses complementarity from (\ref{eq:slackness_r_upper_bound_unconstr}), and substitutes $\ubar{\mu}^l_{r,t}$ as in (\ref{eq:kkt_G_r_l_unconstr}). Third, the complementarity conditions (\ref{eq:slackness_r_lower_bound_unconstr}) and (\ref{eq:slackness_r_upper_bound_unconstr}) are exploited to cancel out terms, while the cycling sum over $G^l_{r,t}$ is shifted to relate the timing to $\lambda^l_{r,t}$. Last, primal feasibility of (\ref{eq:sto_level}) is exploited.

Dividing both sides of Equation~(\ref{eq:derivation_zpr_r_unconstr}) by annual energy output of storage $\sum_t G_{s,t}^{out}$ and rearranging terms renders the storage zero-profit condition in the long-term equilibrium:

\begin{equation}
    LCOS_r = MV_r
\end{equation}

\subsection{Derivation of optimality conditions with binding renewable targets}\label{ssec:derivations_constrained_opt}

When imposing a binding renewable target, the renewable energy constraints (\ref{eq:omega}) and (\ref{eq:theta}) become effective. The \ac{KKT} stationarity conditions (\ref{eq:kkt_G_r_out_theta}) and (\ref{eq:kkt_G_r_out_omega}) refer to models using constraints (\ref{eq:omega}) or (\ref{eq:theta}), respectively, and replace the stationarity condition (\ref{eq:kkt_G_r_out_unconstr}) of the unconstrained model. Similarly, (\ref{eq:kkt_G_r_in_theta}) and (\ref{eq:kkt_G_r_in_omega}) replace (\ref{eq:kkt_G_r_in_unconstr}):



\begin{subequations}
    \begin{align}
        \frac{\partial \mathcal{L}}{\partial G^{out}_{r,t}} &= 0 \Rightarrow o^{out}_r - \lambda_t - \ubar{\mu}^{out}_{r,t} + \bar{\mu}^{out}_{r,t} - (\eta^{out}_r)^{-1} \lambda^l_{r,t} + \frac{\partial \Theta}{\partial G^{out}_{r,t}} \mu_{\theta} = 0 \label{eq:kkt_G_r_out_theta} \\
        \frac{\partial \mathcal{L}}{\partial G^{out}_{r,t}} &= 0 \Rightarrow o^{out}_r - \lambda_t - \ubar{\mu}^{out}_{r,t} + \bar{\mu}^{out}_{r,t} - (\eta^{out}_r)^{-1} \lambda^l_{r,t} - \frac{\partial \Omega}{\partial G^{out}_{r,t}} \mu_{\omega} = 0 \label{eq:kkt_G_r_out_omega}
    \end{align}
\end{subequations}

\begin{subequations}
    \begin{align}
        \frac{\partial \mathcal{L}}{\partial G^{in}_{r,t}} &= 0 \Rightarrow o^{in}_r + \lambda_t - \ubar{\mu}^{in}_{r,t} + \bar{\mu}^{in}_{r,t} + \eta^{in}_r \lambda^l_{r,t} + \frac{\partial \Theta}{\partial G^{in}_{r,t}} \mu_{\theta} = 0 \label{eq:kkt_G_r_in_theta} \\
        \frac{\partial \mathcal{L}}{\partial G^{in}_{r,t}} &= 0 \Rightarrow o^{in}_r + \lambda_t - \ubar{\mu}^{in}_{r,t} + \bar{\mu}^{in}_{r,t} + \eta^{in}_r \lambda^l_{r,t} - \frac{\partial \Omega}{\partial G^{in}_{r,t}} \mu_{\omega} = 0 \label{eq:kkt_G_r_in_omega}
    \end{align}
\end{subequations}

\vspace{0.2cm}

Derivations (\ref{eq:kkt_G_r_out_theta}) and (\ref{eq:kkt_G_r_out_omega}) are alternatives, referring to model configurations based on constraint family (1) and (2) as defined in equation (\ref{eq:theta}), or constraint family (3) and (4) as in equation (\ref{eq:omega}), respectively. The same holds true for (\ref{eq:kkt_G_r_in_theta}) and (\ref{eq:kkt_G_r_in_omega}). The optimality conditions for storage discharging are:

\begin{subequations}
    \begin{align}
        \Theta_{1a} &:\lambda_t = o^{out}_r - \ubar{\mu}^{out}_{r,t} + \bar{\mu}^{out}_{r,t} - (\eta^{out}_r)^{-1} \lambda^l_{r,t} \label{eq:opt_cond_sto_out_1a} \\
        \Theta_{1b} &:\lambda_t = o^{out}_r - \ubar{\mu}^{out}_{r,t} + \bar{\mu}^{out}_{r,t} - (\eta^{out}_r)^{-1} \lambda^l_{r,t} - \phi \mu_{\theta} \\
        \Theta_{1c} &:\lambda_t = o^{out}_r - \ubar{\mu}^{out}_{r,t} + \bar{\mu}^{out}_{r,t} - (\eta^{out}_r)^{-1} \lambda^l_{r,t} - \mu_{\theta}\\
        \Theta_{2a} &:\lambda_t = o^{out}_r - \ubar{\mu}^{out}_{r,t} + \bar{\mu}^{out}_{r,t} - (\eta^{out}_r)^{-1} \lambda^l_{r,t} + \phi \mu_{\theta}\\
        \Theta_{2b} &:\lambda_t = o^{out}_r - \ubar{\mu}^{out}_{r,t} + \bar{\mu}^{out}_{r,t} - (\eta^{out}_r)^{-1} \lambda^l_{r,t} \\
        \Theta_{2c} &:\lambda_t = o^{out}_r - \ubar{\mu}^{out}_{r,t} + \bar{\mu}^{out}_{r,t} - (\eta^{out}_r)^{-1} \lambda^l_{r,t} - (1 -\phi) \mu_{\theta}\\
        \Omega_{3a} &:\lambda_t = o^{out}_r - \ubar{\mu}^{out}_{r,t} + \bar{\mu}^{out}_{r,t} - (\eta^{out}_r)^{-1} \lambda^l_{r,t} + \mu_{\omega}\\
        \Omega_{3b} &:\lambda_t = o^{out}_r - \ubar{\mu}^{out}_{r,t} + \bar{\mu}^{out}_{r,t} - (\eta^{out}_r)^{-1} \lambda^l_{r,t} + (1 - \phi) \mu_{\omega} \\
        \Omega_{3c} &:\lambda_t = o^{out}_r - \ubar{\mu}^{out}_{r,t} + \bar{\mu}^{out}_{r,t} - (\eta^{out}_r)^{-1} \lambda^l_{r,t} \\
        \Omega_{4a} &:\lambda_t = o^{out}_r - \ubar{\mu}^{out}_{r,t} + \bar{\mu}^{out}_{r,t} - (\eta^{out}_r)^{-1} \lambda^l_{r,t} + \phi \mu_{\omega} \\
        \Omega_{4b} &:\lambda_t = o^{out}_r - \ubar{\mu}^{out}_{r,t} + \bar{\mu}^{out}_{r,t} - (\eta^{out}_r)^{-1} \lambda^l_{r,t} \\
        \Omega_{4c} &:\lambda_t = o^{out}_r - \ubar{\mu}^{out}_{r,t} + \bar{\mu}^{out}_{r,t} - (\eta^{out}_r)^{-1} \lambda^l_{r,t} - (1 - \phi) \mu_{\omega} \label{eq:opt_cond_sto_out_4c}
    \end{align}
\end{subequations}

Equivalently, the optimality conditions for storage charging are:

\begin{subequations}
    \begin{align}
        \Theta_{1a} &:\lambda_t = - o^{in}_r + \ubar{\mu}^{in}_{r,t} - \bar{\mu}^{in}_{r,t} - \eta^{in}_r \lambda^l_{r,t} \label{eq:opt_cond_sto_in_1a}\\
        \Theta_{1b} &:\lambda_t = - o^{in}_r + \ubar{\mu}^{in}_{r,t} - \bar{\mu}^{in}_{r,t} - \eta^{in}_r \lambda^l_{r,t} - \phi \mu_{\theta} \\
        \Theta_{1c} &:\lambda_t = - o^{in}_r + \ubar{\mu}^{in}_{r,t} - \bar{\mu}^{in}_{r,t} - \eta^{in}_r \lambda^l_{r,t} - \mu_{\theta} \\
        \Theta_{2a} &:\lambda_t = - o^{in}_r + \ubar{\mu}^{in}_{r,t} - \bar{\mu}^{in}_{r,t} - \eta^{in}_r \lambda^l_{r,t} + \phi \mu_{\theta} \\
        \Theta_{2b} &:\lambda_t = - o^{in}_r + \ubar{\mu}^{in}_{r,t} - \bar{\mu}^{in}_{r,t} - \eta^{in}_r \lambda^l_{r,t} \\
        \Theta_{2c} &:\lambda_t = - o^{in}_r + \ubar{\mu}^{in}_{r,t} - \bar{\mu}^{in}_{r,t} - \eta^{in}_r \lambda^l_{r,t} - (1 - \phi) \mu_{\theta} \\
        \Omega_{3a} &:\lambda_t = - o^{in}_r + \ubar{\mu}^{in}_{r,t} - \bar{\mu}^{in}_{r,t} - \eta^{in}_r \lambda^l_{r,t} + \mu_{\omega} \\
        \Omega_{3b} &:\lambda_t = - o^{in}_r + \ubar{\mu}^{in}_{r,t} - \bar{\mu}^{in}_{r,t} - \eta^{in}_r \lambda^l_{r,t} + (1-\phi) \mu_{\omega} \\
        \Omega_{3c} &:\lambda_t = - o^{in}_r + \ubar{\mu}^{in}_{r,t} - \bar{\mu}^{in}_{r,t} - \eta^{in}_r \lambda^l_{r,t} \\
        \Omega_{4a} &:\lambda_t = - o^{in}_r + \ubar{\mu}^{in}_{r,t} - \bar{\mu}^{in}_{r,t} - \eta^{in}_r \lambda^l_{r,t} + \phi \mu_{\omega} \\
        \Omega_{4b} &:\lambda_t = - o^{in}_r + \ubar{\mu}^{in}_{r,t} - \bar{\mu}^{in}_{r,t} - \eta^{in}_r \lambda^l_{r,t} \\
        \Omega_{4c} &:\lambda_t = - o^{in}_r + \ubar{\mu}^{in}_{r,t} - \bar{\mu}^{in}_{r,t} - \eta^{in}_r \lambda^l_{r,t} - (1 - \phi) \mu_{\omega} \label{eq:opt_cond_sto_in_4c}
    \end{align}
\end{subequations}

\noindent By way of example, the zero-profit rule for storage in the model specification (1c) is as follows:

\begin{equation}\label{eq:derivation_zpr_r_theta1c}
\begin{split}
    \sum_\circ i_r^\circ C_r^\circ &+ \sum_{t,*} o_r^* G_{r,t}^* + \sum_t \lambda_t G_{r,t}^{in}  \\
    & = \sum_{t,\circ} \bar{\mu}^\circ_{r,t} C^\circ_r + \sum_t G_{r,t}^{out} \Big(\lambda_t + \ubar{\mu}^{out}_{r,t} - \bar{\mu}^{out}_{r,t} + \lambda^l_{r,t} (\eta^{out}_r)^{-1} - \frac{\partial \Theta_{1c}}{\partial G_{r,t}^{out}} \mu_{\theta} \Big) \\
    & \quad + \sum_t G_{r,t}^{in} \Big(- \lambda_t + \ubar{\mu}^{in}_{r,t} - \bar{\mu}^{in}_{r,t} - \lambda^l_{r,t} \eta^{in}_r - \frac{\partial \Theta_{1c}}{\partial G_{r,t}^{in}} \mu_{\theta} \Big) + \sum_t \lambda_t G_{r,t}^{in} \\
    & = \sum_{t,\circ} \bar{\mu}^\circ_{r,t} C^\circ_r + \sum_t G_{r,t}^{out} \Big(\lambda_t + \ubar{\mu}^{out}_{r,t} - \bar{\mu}^{out}_{r,t} + \lambda^l_{r,t} (\eta^{out}_r)^{-1} + \mu_{\theta} \Big) \\
    & \quad + \sum_t G_{r,t}^{in} \Big(- \lambda_t + \lambda_t + \ubar{\mu}^{in}_{r,t} - \bar{\mu}^{in}_{r,t} - \lambda^l_{r,t} \eta^{in}_r - \mu_{\theta} \Big) \\
    & = \sum_t \Big( \ubar{\mu}^{out}_{rt} G^{out}_{r,t} + \ubar{\mu}^{in}_{rt} G^{in}_{r,t} + \bar{\mu}^{out}_{rt} (C^{out}_r - G^{out}_{r,t}) + \bar{\mu}^{in}_{rt} (C^{in}_r - G^{in}_{r,t}) \\
    & \quad + G^{out}_{r,t} \big(\lambda_t + \lambda^l_{r,t} (\eta^{out}_r)^{-1} + \mu_{\theta}\big) + G^{in}_{r,t} \big(- \lambda^l_{r,t} \eta^{in}_r - \mu_{\theta}\big)\\
    & \quad + G^l_{r,t} \big( \ubar{\mu}^l_{r,t} + \lambda^l_{r,t} - \lambda^l_{r,t+1} \big) \Big) \\
    &= \sum_t \Big( \lambda_t G_{r,t}^{out} - \mu_{\theta} \big(G_{r,t}^{in} - G_{r,t}^{out}\big) + \ubar{\mu}^l_{r,t} G^l_{r,t} \\
    & \quad + \lambda^l_{r,t} \big(G^l_{r,t} - \eta_r^l G^l_{r,t-1} - \eta^{in}_r G^{in}_{r,t} + (\eta^{out}_r)^{-1} G^{out}_{r,t} \big) \Big)\\
    &= \sum_t \Big( \lambda_t G_{r,t}^{out} - \mu_{\theta} \big(G_{r,t}^{in} - G_{r,t}^{out}\big) \Big)
\end{split}
\end{equation}
\vspace{0.1cm}

The first step employs stationarity conditions (\ref{eq:kkt_C_r_unconstr}), (\ref{eq:kkt_G_r_out_theta}) and (\ref{eq:kkt_G_r_in_theta}). Second, the remaining derivatives are computed. The third step rearranges terms, uses complementarity from (\ref{eq:slackness_r_upper_bound_unconstr}), and substitutes $\ubar{\mu}^l_{r,t}$ as in (\ref{eq:kkt_G_r_l_unconstr}). Fourth, the complementarity conditions (\ref{eq:slackness_r_lower_bound_unconstr}) and (\ref{eq:slackness_r_upper_bound_unconstr}) are exploited to cancel out terms, while the cycling sum over $G^l_{r,t}$ is shifted to relate the timing to $\lambda^l_{r,t}$. Last, primal feasibility of (\ref{eq:sto_level}) is exploited.

Dividing both sides of equation (\ref{eq:derivation_zpr_r_theta1c}) by the storage unit's annual energy output $\sum_t G_{s,t}^{out}$ and rearranging terms renders its long-term equilibrium condition, as in Table~\ref{tab:zero_profit_r}:

\begin{equation}
    LCOS_r + \mu_{\theta} NSL_r = MV_r
\end{equation}

The mathematical proofs zero-profit conditions of all other investigated model specifications follow the same procedure and are available upon request.

\subsection{Specifics of numerical results of constraint family (3)}\label{ssec:con_family_3}

\begin{figure}[htbp]
\centering
\noindent\includegraphics[width=\textwidth, keepaspectratio]{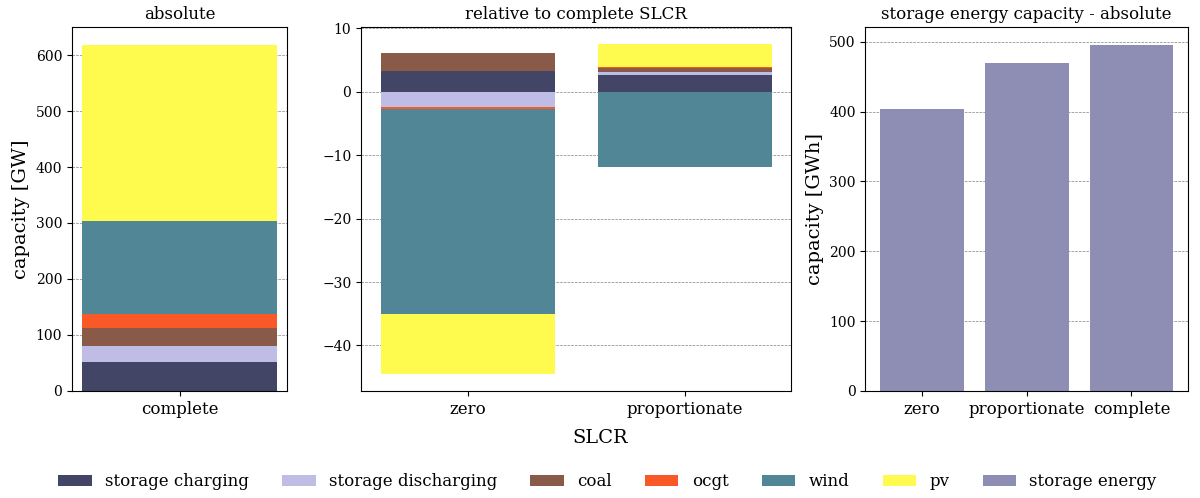}
\caption{Total capacity per technology at a renewable target of 80\% in absolute numbers for model specifications with complete \ac{SLCR} (left panel), and the deviation from the complete \ac{SLCR} case for models with zero and proportionate \ac{SLCR} for constraint families (1), (2), and~(4) (middle panel), and for constraint family~(3) (right panel).}
\label{fig:cap_annual}
\end{figure}

\begin{figure}[htbp]
\centering
\noindent\includegraphics[width=\textwidth, keepaspectratio]{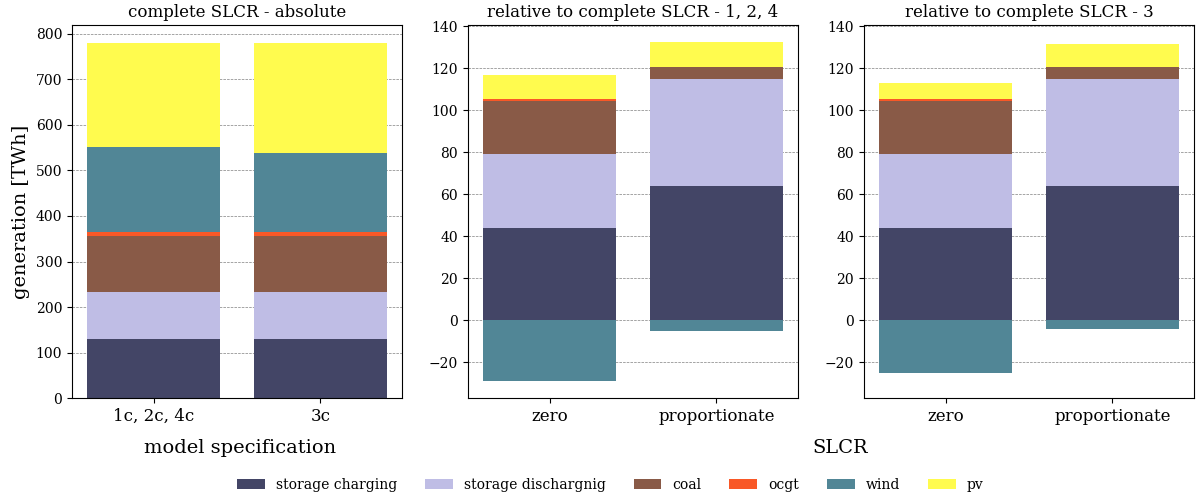}
\caption{Total generation per technology at a renewable target of 80\% in absolute numbers for model specifications with complete \ac{SLCR} (left panel), and the deviation from the complete \ac{SLCR} case for models with zero and proportionate \ac{SLCR} for constraint families (1), (2), and~(4) (middle panel), and for constraint family~(3) (right panel).}
\label{fig:gen_annual}
\end{figure}

Dispatch from \ac{VRE} is more similar between different \ac{SLCR} levels within constraint family (3) models compared to the other constraint families (Figure \ref{fig:gen_annual}). Furthermore, models based on constraint family (3) slightly favor \ac{PV} over wind compared to all other models using other constraint families. In contrast, optimal investment decisions coincide across all constraint families, and vary only with respect to the \ac{SLCR} level (Figure~\ref{fig:cap_annual}).

\subsection{Factor separation: unintended storage cycling and ambition levels of the renewable constraint}\label{ssec:factor_separation}

The differences in optimal capacity and dispatch decisions presented in Figure~\ref{fig:gen_cap} are caused by variations of the \ac{SLCR} level in the renewable energy constraint. Two overlapping factors drive results: (i) unintended storage cycling, and (ii) different ambition levels regarding the required minimum renewable penetration. The latter increases as more storage losses have to be covered by renewable energy. As explained in Section \ref{ssec:res_con_formulation}, models with zero \ac{SLCR} require the lowest use of \ac{VRE}, followed by models with proportionate \ac{SLCR}, and such with complete \ac{SLCR}, which require the highest use of \ac{VRE}. We disentangle the impacts of both factors by carrying out additional model runs with adjusted renewable energy targets. 

First, the effects of unintended storage cycling can be isolated when contrasting a model with incomplete \ac{SLCR} to a setting with complete \ac{SLCR}. To rule out the impacts of different ambition levels regarding the renewable target, the realized renewable share of both settings needs to coincide when reported according to the renewable share formula of the model with incomplete \ac{SLCR}. 

In our stylized setting, a complete \ac{SLCR} model with a 77.0\% renewable target achieves the same renewable penetration as a zero \ac{SLCR} model with an 80.0\% renewable target, when reporting both renewable penetration levels according to the zero \ac{SLCR} specification. Likewise, the renewable penetration of a model with proportionate \ac{SLCR} and an 80\% target matches that of a model with a complete \ac{SLCR} and a 79.3\% target, when reported according to the proportionate \ac{SLCR} formula. Note that determining such matching renewable targets for complete \ac{SLCR} settings, in which unintended storage cycling never occurs, may require a time-consuming try-and-error approach, which needs to be carried out for each individual model parameterization. 

Second, the impact of different ambition levels of the required renewable energy share can be determined by comparing complete \ac{SLCR} specifications with different renewable shares.
In our stylized setting, contrasting the additional specifications with complete \ac{SLCR} and a 77.0\% or 79.3\% renewable target to a complete \ac{SLCR} model with 80\% allows for separating the effects of different ambition levels from unintended storage cycling for zero and proportionate \ac{SLCR} settings with an 80\% renewable penetration, respectively.

\begin{figure}[htbp]%
\centering
\subfloat[Installed generation capacity including storage charging and discharging.\label{fig:cap_fs}]
    {{\includegraphics[width=0.48\textwidth]{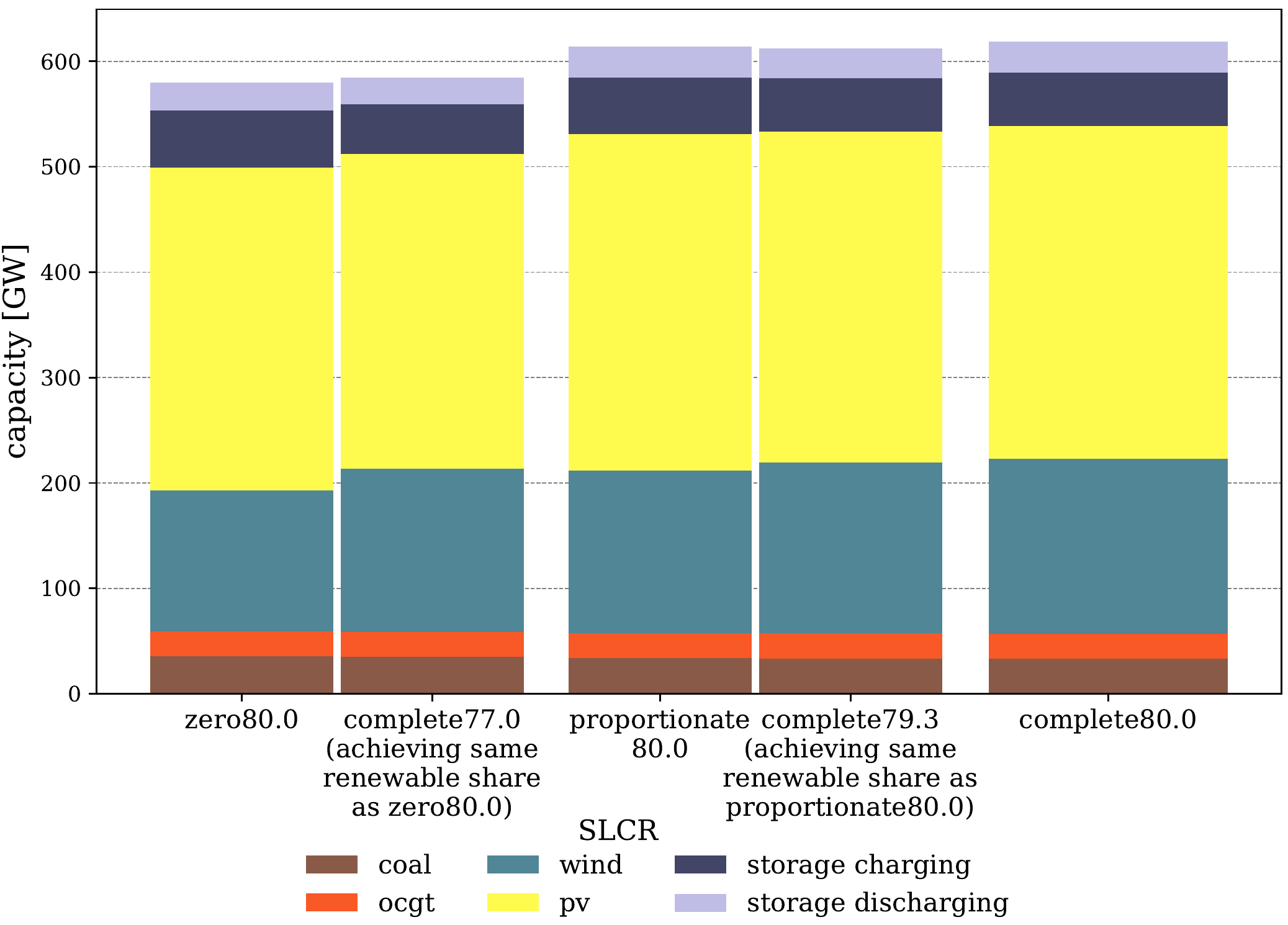}}}
\quad
\subfloat[Installed storage energy capacity.\label{fig:cap_sto_e_fs}]
    {{\includegraphics[width=0.48\textwidth]{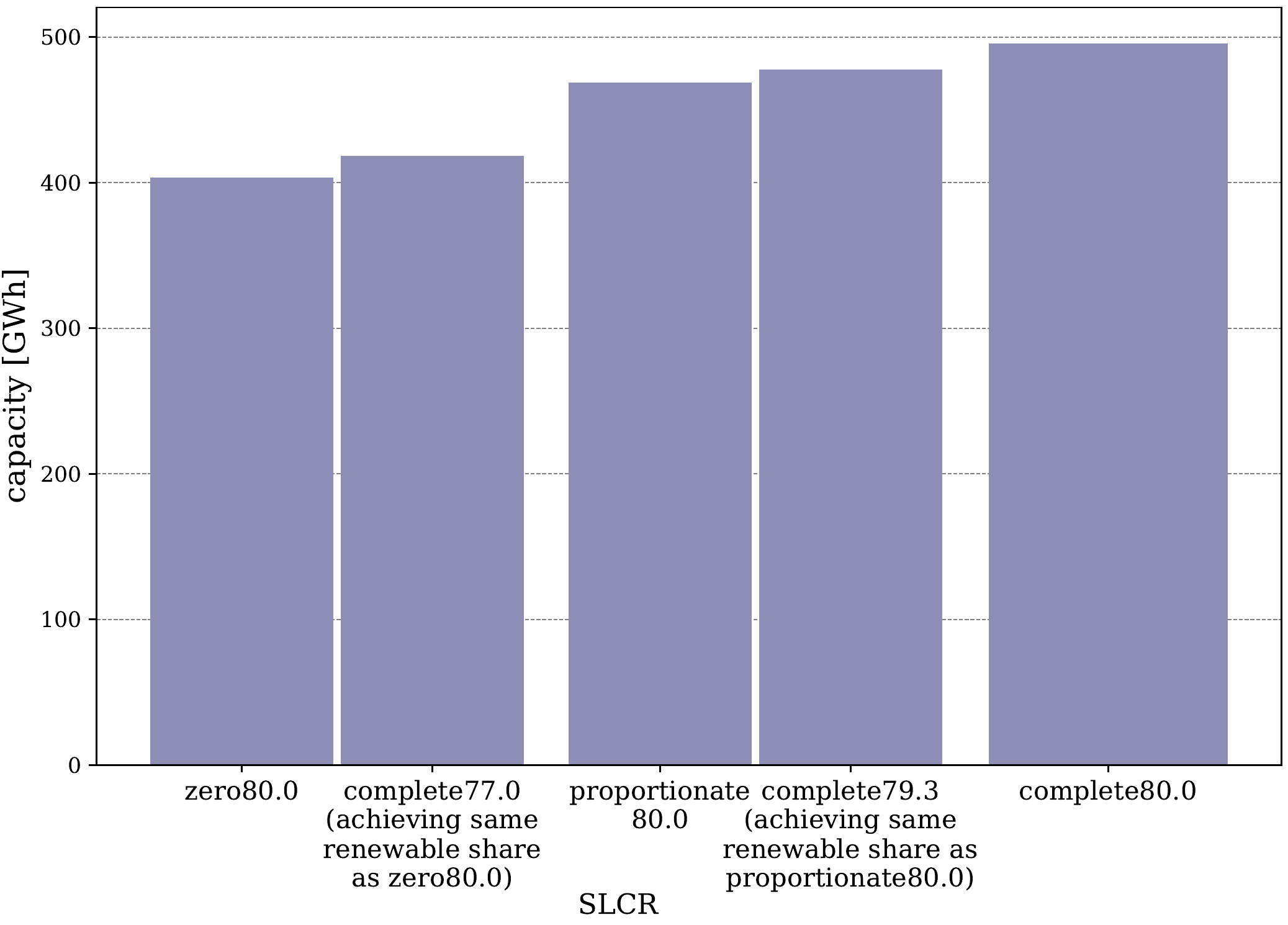}}}\\
\subfloat[Generation including storage charging (negative part of ordinate) and discharging.\label{fig:gen_fs}]
    {{\includegraphics[width=0.48\textwidth]{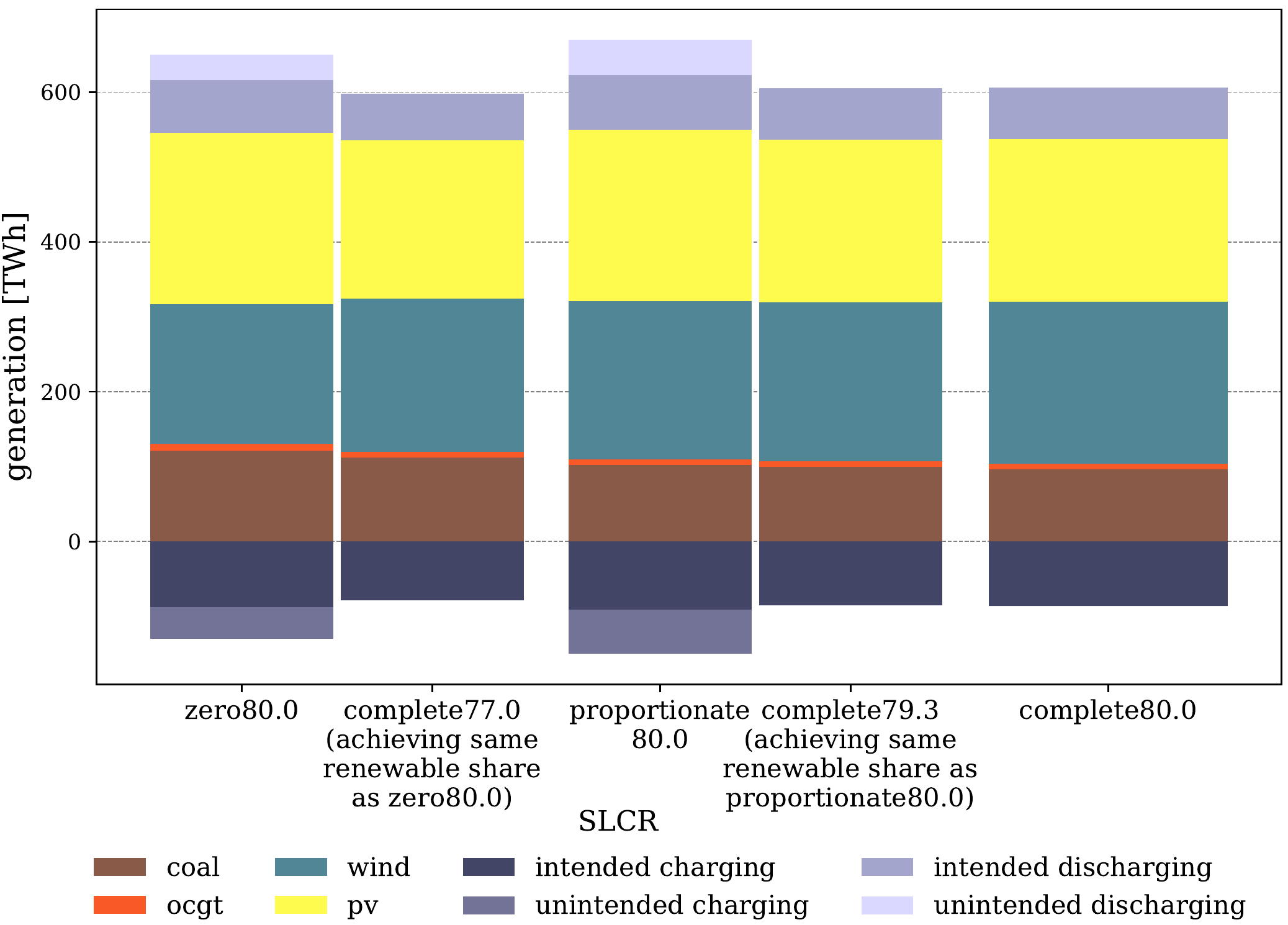}}}
\quad
\subfloat[Storage losses and curtailment.\label{fig:sto_loss_cu_fs}]
    {{\includegraphics[width=0.48\textwidth]{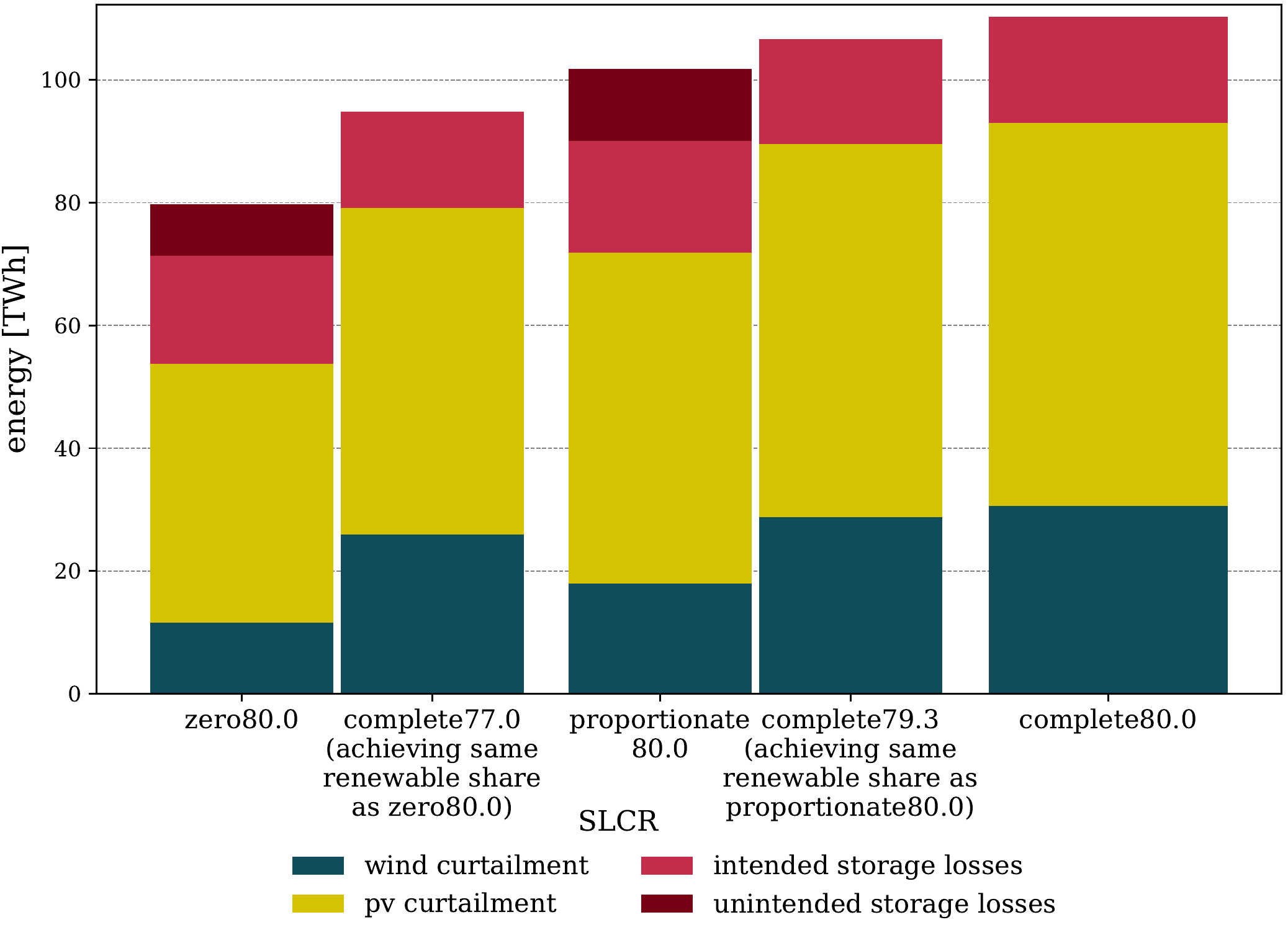}}}
\caption{Installed capacity, annual generation, curtailment, and storage losses per technology, including factor separation model runs. The numbers in the labels refer to renewable targets specified in respective models.}
\label{fig:fac_sep}%
\end{figure}

Figure~\ref{fig:fac_sep} illustrates the outcomes of the two additional model specifications with adjusted renewable energy targets (77.0\% and 79.3\%), which allow for disentangling the effects of both factors. Comparing the cases with complete \ac{SLCR}, i.e., models without unintended storage cycling, it can be seen that total \ac{VRE} capacity decreases with lower ambition levels (Figure~\ref{fig:cap_fs}). 
This decrease is the main driver for lower storage energy capacity needs presented in Section~\ref{ssec:results_dispatch_investment} (Figure~\ref{fig:cap_sto_e_fs}).

Comparing the zero and proportionate \ac{SLCR} models with their counterparts with complete \ac{SLCR} and adjusted renewable shares (i.e.,~comparing columns 1 and 2, as well as columns 3 and 4), it can be seen that unintended storage cycling causes an additional \ac{VRE} capacity effect, which is somewhat lower than the one attributed to the ambition level (Figure~\ref{fig:cap_fs}). There is also an impact on the optimal \ac{VRE} portfolio. In total, the need for \ac{VRE} capacity decreases due to a conversion of curtailment into additional \ac{VRE} generation that contributes to renewable energy constraint. However, this decrease applies solely to wind, while \ac{PV} capacity even increases. This is because \ac{LCOE} of wind power (not accounting for its generation profile) are higher than those of PV in the model parameterization used here. Generation from \ac{VRE} is similarly affected (Figure~\ref{fig:gen_fs}). In models with proportionate \ac{SLCR}, there is a disproportionate shift from wind to \ac{PV}, and the total \ac{VRE} generation even increases. The latter is because only a fraction of unintended storage losses contribute to achieving the renewable energy target. Note that the increase in \ac{VRE} generation between column 4 to column 3 does not incur any costs, as it goes along with lower renewable curtailment. The portfolio effect smooths \ac{VRE} generation, as \ac{PV} has a more regular diurnal generation profile than wind power. This contributes to a slightly decreasing need for storage energy capacity (Figure~\ref{fig:cap_sto_e_fs}).

The need for storage charging and discharging capacity slightly decreases as the renewable ambition level declines (compare columns 2 and 5, as well as columns 4 and 5 in Figure~\ref{fig:cap_fs}). This is most pronounced in models with zero \ac{SLCR}, and driven by the \ac{VRE} capacity effect of the lower ambition level. Intended storage use hardly changes with lower ambition levels (Figure~\ref{fig:gen_fs}). In contrast, optimal charging and discharging capacity increases in settings with unintended storage cycling, as it helps to extend the unintended storage cycling potential. Here, additional investment costs for storage capacities are lower than savings due to the \ac{VRE} capacity and portfolio effect. Intended storage use and storage losses slightly increases with unintended storage cycling driven by these storage capacity effects. 

The increase in generation from coal plants in models with incomplete \ac{SLCR} compared to settings with complete \ac{SLCR} can be attributed to both the reduced ambition level of the respective renewable constraint and unintended storage cycling, with the former slightly more effective than the latter (Figure~\ref{fig:gen_fs}). 

Curtailment reduction is primarily caused by the conversion of curtailment into unintended storage losses (compare columns 1 and 2, as well as 3 and 4 in Figure~\ref{fig:sto_loss_cu_fs}). Another driver is the reduction of optimal \ac{VRE} capacity caused by a lower ambition level of the renewable target, which leads to lower \ac{VRE} surplus generation. However, this effect plays less of a role (compare decrease from column 5 to columns 2 and 4 with decrease caused by unintended storage cycling in Figure~\ref{fig:sto_loss_cu_fs}).

\subsection{An example for renewable energy constraints for sector coupling options}\label{ssec:usc_remedy_sector_soupling}

In energy models covering the power and transport sectors that include \ac{BEV} with a discharge option (vehicle-to-grid), unintended storage cycling may also arise in vehicle batteries. To prevent this from happening, any possible loss arising from energy cycling not only of stationary electricity storage, but also of \ac{BEV} (via re-conversion) has to be accounted for in the loss term of the renewable energy constraint. Unavoidable losses associated with charging energy to supply the electrical demand for driving cannot be exploited for additional energy losses, as long as this demand is exogenous to the model. Hence, they can be excluded from loss term. Be $v$ a member of the set of all \ac{BEV}, $BEV^{in}_{v,t}$ and $BEV^{out}_{v,t}$ charging and discharging from and back into the grid of each \ac{BEV} in period $t$, $BEV^l_t$ the battery level, $d^{BEV}_{v,t}$ the required energy for driving, and $\eta^{BEV}_{in}$ the battery's charging efficiency. Further, to avoid free lunch, we assume that the battery level in the first and last hour of the year need to be equal $BEV^l_1 = BEV^l_{T}$. We recommend to supplement the loss term as displayed in Table \ref{tab:res_shares} with the following concise term: 

\begin{equation}\label{eq:loss_bev1}
    losses^{BEV} = \sum\limits_{v, t} \big(BEV^{in}_{v,t} - BEV^{out}_{v,t} - d^{BEV}_{v,t}/\eta^{BEV}_{in} \big)
\end{equation}

\vspace{0.3cm}

Losses comprise charged energy reduced by discharged energy and energy for driving, corrected by conversion losses from charging. If the battery level in the first and last period is allowed to diverge, the loss term needs to be corrected by the absolute change in storage level across the entire optimization horizon:

\begin{equation}
    losses^{BEV} = \sum\limits_{v, t} \big(BEV^{in}_{v,t} - BEV^{out}_{v,t} - d^{BEV}_{v,t}/\eta^{BEV}_{in} - BEV^l_T + BEV^l_1 \big)
\end{equation}

\vspace{0.3cm}

For the loss term (\ref{eq:loss_bev1}) without free lunch, the minimal renewable energy constraint (2c) with complete \ac{SLCR} referenced to total generation, which avoids unintended energy cycling in electricity storage and \ac{BEV}, would be: 

\begin{equation}
    \sum\limits_{s \in \mathcal{R}, t} G_{s,t} \geq \phi \sum\limits_{s,t} G_{s,t} + (1-\phi) \Big(\sum\limits_{r,t} (G^{in}_{r,t} - G^{out}_{r,t}) + \sum\limits_{v, t} (BEV^{in}_{v,t} - BEV^{out}_{v,t} - d^{BEV}_{v,t}/\eta^{BEV}_{in}) \Big)
\end{equation}

If the use of multiple vehicle technologies, e.g., \ac{BEV}, electric vehicle with a hydrogen-fuel fuel cell, or vehicles with an internal combustion engine, is endogenously determined, conversion losses associated with demand for driving need to added to the loss term of the renewable energy constraint, too.

\end{document}